\documentclass[preprintnumbers,twocolumn, superscriptaddress,nofootinbib,aps,prd,floatfix]{revtex4}
\pdfoutput=1

\usepackage{multirow,array}
\usepackage{amsmath,amssymb}
\usepackage{graphicx}
\usepackage{color}
\usepackage{url}
\usepackage{feynmp}
\usepackage{slashed}
\usepackage{multirow}
\usepackage{footmisc}
\usepackage{hyperref}
\usepackage[scientific-notation=true]{siunitx}
\usepackage{dsfont}
\usepackage{ulem}
\usepackage{cancel}

\hypersetup{
  pdfauthor={K.S. Babu, Dorival Gon\c{c}alves, Ahmed Ismail},
  pdftitle={Lepton Number Violation at the LHC},
}
\begin{document}
\def\contentsname{{\normalsize Content}}
\def\tablename{Table}
\def\figurename{Figure}

\def\bdrs{\text{BDRS}}
\def\pveto{P_\text{veto}}
\def\nj{n_\text{jets}}
\def\meff{m_\text{eff}}
\def\ptmin{p_T^\text{min}}
\def\gtot{\Gamma_\text{tot}}
\def\as{\alpha_s}
\def\az{\alpha_0}
\def\gz{g_0}
\def\w{\vec{w}}
\def\sdag{\Sigma^{\dag}}
\def\s{\Sigma}
\newcommand{\psib}{\overline{\psi}}
\newcommand{\Psib}{\overline{\Psi}}
\newcommand\one{\leavevmode\hbox{\small1\normalsize\kern-.33em1}}
\newcommand{\Mpl}{M_\mathrm{Pl}}
\newcommand{\p}{\partial}
\newcommand{\mat}{\mathcal{M}}
\newcommand{\lag}{\mathcal{L}}
\newcommand{\ord}{\mathcal{O}}
\newcommand{\ope}{\mathcal{O}}
\newcommand{\qqquad}{\qquad \qquad}
\newcommand{\qqqquad}{\qquad \qquad \qquad}

\newcommand{\qb}{\bar{q}}
\newcommand{\matx}{|\mathcal{M}|^2}
\newcommand{\really}{\stackrel{!}{=}}
\newcommand{\msbar}{\overline{\text{MS}}}
\newcommand{\qns}{f_q^\text{NS}}
\newcommand{\lqcd}{\Lambda_\text{QCD}}
\newcommand{\met}{E_T^{\text{miss}}}
\newcommand{\pmiss}{\slashchar{\vec{p}}_T}

\newcommand{\sq}{\tilde{q}}
\newcommand{\go}{\tilde{g}}
\newcommand{\st}[1]{\tilde{t}_{#1}}
\newcommand{\stb}[1]{\tilde{t}_{#1}^*}
\newcommand{\nz}[1]{\tilde{\chi}_{#1}^0}
\newcommand{\cp}[1]{\tilde{\chi}_{#1}^+}
\newcommand{\CP}{CP}

\providecommand{\mg}{m_{\tilde{g}}}
\providecommand{\mst}[1]{m_{\tilde{t}_{#1}}}
\newcommand{\msn}[1]{m_{\tilde{\nu}_{#1}}}
\newcommand{\mch}[1]{m_{\tilde{\chi}^+_{#1}}}
\newcommand{\mne}[1]{m_{\tilde{\chi}^0_{#1}}}
\newcommand{\msb}[1]{m_{\tilde{b}_{#1}}}
\newcommand{\vsm}{\ensuremath{v_{\rm SM}}}

\newcommand{\mev}{{\ensuremath\rm MeV}}
\newcommand{\gev}{{\ensuremath\rm GeV}}
\newcommand{\tev}{{\ensuremath\rm TeV}}
\newcommand{\sign}{{\ensuremath\rm sign}}
\newcommand{\iab}{{\ensuremath\rm ab^{-1}}}
\newcommand{\ifb}{{\ensuremath\rm fb^{-1}}}
\newcommand{\ipb}{{\ensuremath\rm pb^{-1}}}

\def\slashchar#1{\setbox0=\hbox{$#1$}           
   \dimen0=\wd0                                 
   \setbox1=\hbox{/} \dimen1=\wd1               
   \ifdim\dimen0>\dimen1                        
      \rlap{\hbox to \dimen0{\hfil/\hfil}}      
      #1                                        
   \else                                        
      \rlap{\hbox to \dimen1{\hfil$#1$\hfil}}   
      /                                         
   \fi}
\newcommand{\dslash}{\slashchar{\partial}}
\newcommand{\Dslash}{\slashchar{D}}

\newcommand{\eg}{\textsl{e.g.}\;}
\newcommand{\ie}{\textsl{i.e.}\;}
\newcommand{\etal}{\textsl{et al}\;}

\setlength{\floatsep}{0pt}
\setcounter{topnumber}{1}
\setcounter{bottomnumber}{1}
\setcounter{totalnumber}{1}
\renewcommand{\topfraction}{1.0}
\renewcommand{\bottomfraction}{1.0}
\renewcommand{\textfraction}{0.0}
\renewcommand{\thefootnote}{\fnsymbol{footnote}}

\newcommand{\rig}{\rightarrow}
\newcommand{\lrig}{\longrightarrow}
\renewcommand{\d}{{\mathrm{d}}}
\newcommand{\be}{\begin{eqnarray*}}
\newcommand{\ee}{\end{eqnarray*}}
\newcommand{\gl}[1]{(\ref{#1})}
\newcommand{\ta}[2]{ \frac{ {\mathrm{d}} #1 } {{\mathrm{d}} #2}}
\newcommand{\bee}{\begin{eqnarray}}
\newcommand{\eee}{\end{eqnarray}}
\newcommand{\beeq}{\begin{equation}}
\newcommand{\eeeq}{\end{equation}}
\newcommand{\mc}{\mathcal}
\newcommand{\mr}{\mathrm}
\newcommand{\ep}{\varepsilon}
\newcommand{\emt}{$\times 10^{-3}$}
\newcommand{\emfo}{$\times 10^{-4}$}
\newcommand{\emfi}{$\times 10^{-5}$}

\newcommand{\revision}[1]{{\bf{}#1}}

\newcommand{\hzero}{h^0}
\newcommand{\Hzero}{H^0}
\newcommand{\Azero}{A^0}
\newcommand{\PHiggs}{H}
\newcommand{\PW}{W}
\newcommand{\PZ}{Z}

\newcommand{\sw}{\ensuremath{s_w}}
\newcommand{\cw}{\ensuremath{c_w}}
\newcommand{\swd}{\ensuremath{s^2_w}}
\newcommand{\cwd}{\ensuremath{c^2_w}}

\newcommand{\mhhd}{\ensuremath{m^2_{\Hzero}}}
\newcommand{\mhh}{\ensuremath{m_{\Hzero}}}
\newcommand{\mlhd}{\ensuremath{m^2_{\hzero}}}
\newcommand{\Mlh}{\ensuremath{m_{\hzero}}}
\newcommand{\mad}{\ensuremath{m^2_{\Azero}}}
\newcommand{\mhpd}{\ensuremath{m^2_{\PHiggs^{\pm}}}}
\newcommand{\mhp}{\ensuremath{m_{\PHiggs^{\pm}}}}

 \newcommand{\sa}{\ensuremath{\sin\alpha}}
 \newcommand{\ca}{\ensuremath{\cos\alpha}}
 \newcommand{\cad}{\ensuremath{\cos^2\alpha}}
 \newcommand{\sad}{\ensuremath{\sin^2\alpha}}
 \newcommand{\sbd}{\ensuremath{\sin^2\beta}}
 \newcommand{\cbd}{\ensuremath{\cos^2\beta}}
 \newcommand{\cb}{\ensuremath{\cos\beta}}
 \renewcommand{\sb}{\ensuremath{\sin\beta}}
 \newcommand{\tanbd}{\ensuremath{\tan^2\beta}}
 \newcommand{\cotbd}{\ensuremath{\cot^2\beta}}
 \newcommand{\tanb}{\ensuremath{\tan\beta}}
 \newcommand{\tb}{\ensuremath{\tan\beta}}
 \newcommand{\cotb}{\ensuremath{\cot\beta}}



\title{Probing Lepton Number Violation and Majorana Nature of Neutrinos at the LHC}

\author{K.S. Babu}
\author{Rahool K. Barman}
\author{Dorival Gon\c{c}alves}
\author{Ahmed Ismail}
\affiliation{Department of Physics, Oklahoma State University, Stillwater, OK, 74078, USA}

\begin{abstract}
  \noindent
Observation of lepton number ($L$) violation by two units at colliders would provide evidence for the Majorana nature of neutrinos. We study signals of $L$-violation in the context of two popular models of neutrino masses, the type-II seesaw model and the Zee model, wherein small neutrino masses arise at the tree-level and one-loop level, respectively. We focus on $L$-violation signals at the LHC  arising through the process $pp \to \ell^{\pm}\ell^{\prime \pm}$ + jets within these frameworks. We obtain sensitivity to  $L$-violation in the type-II seesaw model for triplet scalar masses up to 700 GeV and in the Zee model  for charged scalar masses up to 4.8 TeV at the high-luminosity LHC with an integrated luminosity of $3~\text{ab}^{-1}$. 

\end{abstract}

\maketitle

\section{Introduction}
\label{sec:intro}

The existence of small but non-zero neutrino masses, as implied by the neutrino oscillation experiments, is clear evidence for physics beyond the Standard Model (SM). While it is possible that neutrinos are Dirac particles with their masses arising from extremely small Yukawa couplings involving right-handed neutrinos, a more natural scenario would be to introduce their masses via the dimension-five Weinberg operator \cite{Weinberg:1979sa}. In this case lepton number is not conserved and neutrinos are Majorana particles. 
The minimal UV complete possibilities realizing this scheme at the tree-level are the type-I~\cite{Minkowski:1977sc,GellMann:1980vs,Glashow:1979nm,Yanagida:1979as,Mohapatra:1979ia}, type-II~\cite{Magg:1980ut, Schechter:1980gr, Cheng:1980qt, Mohapatra:1980yp}, and type-III~\cite{Foot:1988aq} seesaw models, that extend the SM respectively with right-handed neutrinos, a complex scalar triplet, and fermionic triplets \cite{Ma:1998dn}. 
Alternatively, small Majorana neutrino masses may be induced as quantum corrections arising through loop diagrams. In such models, the scale of new physics can be naturally low, since the neutrino mass is suppressed by a loop factor as well as by charged fermion masses. The  Zee model is the simplest extension that falls in this category, which generates neutrino masses radiatively at one-loop~\cite{Zee:1980ai}. Other model frameworks have been proposed to generate neutrino masses at one-loop~\cite{Hall:1983id},  two-loops~\cite{Zee:1985id, Babu:1988ki}, and  three-loops~\cite{Krauss:2002px}, see Refs.~\cite{Cai:2017mow,Babu:2019mfe} for recent reviews, updates and references.

A Majorana mass term for the neutrino would explicitly break lepton number ($L$) by two units ($|\Delta L| = 2$). Establishing the Majorana nature of the neutrino  would have profound impact in our understanding of the cosmos, since the same interactions can lead to baryon-antibaryon asymmetry of the universe via leptogenesis~\cite{Fukugita:1986hr}. The observation of neutrinoless double decay $(0\nu\beta\beta$-decay)~\cite{PhysRev.56.1184} of atomic nuclei would provide direct evidence for lepton number violation (LNV) by two units, which is being explored experimentally with increased sensitivity (for a recent review see Ref.~\cite{Rodejohann:2011mu}).
At high energy scales, the LHC provides another interesting probe for LNV, which is the focus of this paper. Since  lepton number (LN) is zero in the initial state in $pp$ collisions, it is possible to infer LNV at the LHC by observing final states with non-zero LN. This requires signatures with an excess of leptons or anti-leptons in the final state carrying non-zero lepton number. 
The classic LNV signature at  colliders is the final states with same-sign dilepton  plus jets,  $pp \rightarrow \ell^\pm \ell'^\pm$ + jets, first proposed by Keung and Senjanovic~\cite{Keung:1983uu} in the context of left-right symmetric models (LRSM)~\cite{Pati:1974yy, Mohapatra:1974hk, Mohapatra:1974gc, Mohapatra:1979ia, Mohapatra:1980yp}. Here $pp$ collision would produce a right-handed Majorana neutrino $N$ and a charged lepton $\ell^\pm$ via $s$-channel exchange of a heavy $W_R^\pm$ gauge boson, with $N$ decaying into $\ell'^\pm$ + jets by virtue of its Majorana nature.  The resulting process, $pp \rightarrow \ell^\pm \ell'^\pm$ + jets, clearly shows signs of $L$-violation by two units. Prospects for LNV signatures at the LHC in the LRSM  have been extensively explored~\cite{Ferrari:2000sp, Gninenko:2006br, Maiezza:2010ic, Nemevsek:2011hz, Chen:2011hc, Chakrabortty:2012pp, Aguilar-Saavedra:2012grq, Han:2012vk, Chen:2013foz, Dev:2013wba, Dutta:2014dba, Gluza:2015goa, Ng:2015hba, Maiezza:2015lza, Deppisch:2015qwa, Degrande:2016aje, Dev:2016dja, Roitgrund:2017byx, Nemevsek:2018bbt}.

Several authors have investigated LNV signatures within the general framework of type-I seesaw mechanism~\cite{Dicus:1991fk, Datta:1993nm, Ali:2001gsa,Han:2006ip, Kersten:2007vk,delAguila:2007qnc, Atre:2009rg, Dev:2013wba, Alva:2014gxa, Deppisch:2015qwa, Das:2015toa, Ng:2015hba, Degrande:2016aje, Drewes:2019byd, Fuks:2020att}. Signals analogous to the Keung-Senjanovic process can be realized here as well, with $s$-channel exchange of $W^\pm$ gauge boson producing $N+ \ell^\pm$ pair.  However, the production cross-section for this process is suppressed by the square of the $\nu-N$ mixing angle $\theta_{\nu N}$ which is highly constrained by neutrino mass.  In a one-generation model, this mixing is given by $\theta_{\nu N} \simeq \sqrt{m_\nu/M_N} \sim 10^{-6}$ (for $M_N \sim 100$ GeV), leading to unobservable $L$-violation signals.  With three families of active neutrinos mixing with three right-handed neutrinos, the intricate connection between the active-sterile mixing and the neutrino mass can be evaded by cancellation or by approximate symmetries. However, it has been shown in Ref.~\cite{Kersten:2007vk} that in this case lepton number turns out to be nearly conserved, suppressing $L$-violation signals at the LHC. Typically, the studies of $L$-violation in the context of type-I seesaw model take a phenomenological approach and treat the $\nu-N$ mixing to be independent of the neutrino masses.  

The phenomenology of type-II seesaw model at colliders has been extensively studied in Refs.~\cite{Akeroyd:2007zv,Perez:2008zc,FileviezPerez:2008jbu,Melfo:2011nx}, which are crucial in testing several aspects of the neutrino mass generation mechanism. To our knowledge, a lepton number violating signal has not been shown to be observable at colliders in this framework.\footnote{For instance, there are already stringent limits on doubly charged scalar masses from both ATLAS and CMS, where the channel $pp\to \delta^{\pm\pm}\delta^{\mp\mp}\to \ell^\pm\ell^\pm \ell^\mp\ell^\mp$ plays a leading role~\cite{Aaboud:2017qph, CMS:2017pet, ATLAS:2022pbd}. Whereas the observation of this standard type-II seesaw signal would be a clear evidence of new physics, it does not warrant a sign for LNV as this final state has null lepton number.}

There have been studies of $L$-violation in a general Higgs triplet model in the decay of top quark~\cite{Quintero:2012jy} and in effective field theory approach~\cite{Fuks:2020zbm, Aoki:2020til, Harz:2021psp, Graesser:2022nkv}, which are however not directly tied to the neutrino masses. The complementarity between $0\nu\beta\beta$-decay and LNV searches at the LHC in the same-sign dilepton plus jets channel has also been explored in  simplified model frameworks~\cite{Helo:2013dla,Helo:2013ika,delAguila:2013yaa,delAguila:2013mia,Peng:2015haa,Harz:2021psp, Graesser:2022nkv}. $L$-violation signals at the LHC in explicit neutrino mass models arising from $d=7$ operators has been studied in Ref.~\cite{Cepedello:2017lyo}, and in a colored scalar extension in Ref.~\cite{Carquin:2019xiz}. $L$-violation in Higgs boson decay has been studied in Ref.~\cite{Maiezza:2015lza} in the context of left-right symmetric models.

In the present study, we investigate LNV signatures at the LHC in two popular models that generate Majorana masses for neutrinos at tree-level and one-loop level, respectively, the type-II seesaw model~\cite{Magg:1980ut, Schechter:1980gr, Cheng:1980qt, Mohapatra:1980yp} and the Zee model~\cite{Zee:1980ai}. The masses of new particles in both frameworks can be $\mathcal{O}(\mathrm{TeV})$, well within reach of the LHC. We study their  sensitivity to LNV through the same-sign dilepton plus jets signature, $pp\to \ell^{\pm}\ell'^{\pm}+$~jets,  at the high-luminosity LHC (HL-LHC), where $\ell=e,\mu$. Whereas new physics searches have been extensively explored in the literature for the type-II seesaw model~\cite{Akeroyd:2007zv,Perez:2008zc,FileviezPerez:2008jbu,Melfo:2011nx} and the Zee model~\cite{AristizabalSierra:2006ri,Cai:2017mow,Herrero-Garcia:2017xdu,Babu:2019mfe,Babu:2020ivd,Barman:2021xeq}, the present paper derives, for the first time, the sensitivity to {\it lepton number violation} in these two popular frameworks. We also draw some comparisons on the constraints between the considered search channel and the standard ones with null lepton number in the final states.

Observation of $\Delta L = 2$ signal at the LHC via the process $pp \rightarrow \ell^\pm \ell'^\pm$ + jets would imply that neutrinos are Majorana particles.  This inference is possible by virtue of a black box theorem discussed in Sec.~\ref{sec:bbt} which was originally applied to neutrinoless double beta decay~\cite{Schechter:1981bd}, which we extend to collider signals. Unlike $0\nu\beta\beta$ signals which applies only to electron flavor, at colliders any flavor of leptons with same sign would result in the conclusion that neutrinos are Majorana particles.

The rest of this paper is organized as follows. We discuss the correspondence between LNV at colliders and  Majorana masses for neutrinos in Sec.~\ref{sec:bbt}. In Section~\ref{sec:type-II}, we study the projected HL-LHC sensitivity for LNV with same-sign dilepton plus jets in the type-II seesaw model. Section~\ref{sec:zee-model} provides an analogous interpretation for the Zee model. We conclude in Section~\ref{sec:summary}.


\section{Black box theorem at Colliders}
\label{sec:bbt}

\begin{figure}[!b]
    \centering
    \includegraphics[width=0.42\textwidth]{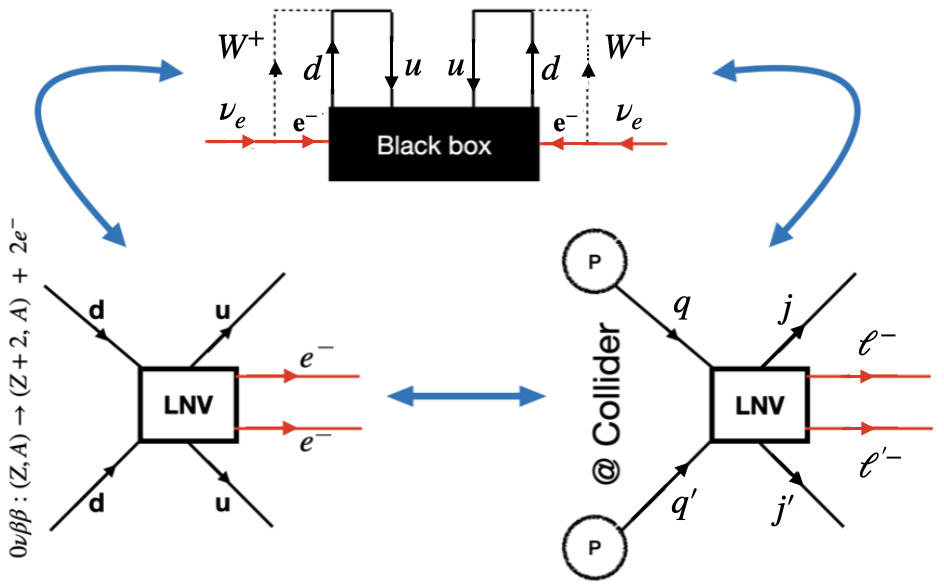}
    \caption{Diagram illustrating the correspondence among Majorana mass generation for neutrinos (top), $0\nu\beta\beta$ decay (bottom-left), and lepton number violating signatures  at the LHC (bottom-right). In the botom-right diagram $\ell^-\ell'^-$ in the final state can be replaced by $\ell^+ \ell'^+$ as well.  All the  diagrams depicted violate lepton number by two units, $|\Delta L|=2$.}
    \label{fig:black_box}
\end{figure}

The \textit{black box theorem}~\cite{Schechter:1981bd} establishes a direct correlation between the observation of $0\nu\beta\beta$ decay and  Majorana masses for the neutrinos. A cartoon representation of the $0\nu\beta\beta$ decay contribution to the neutrino Majorana mass is illustrated in Fig.~\ref{fig:black_box}. This theorem ensures that the LNV interactions, which imply non-zero rates for $0\nu\beta\beta$ decay, will also induce non-zero Majorana mass for  neutrinos, at least at four-loop, irrespective of the underlying new physics model generating LNV. Although the contribution from $0\nu\beta\beta$ decay to the neutrino masses can be extremely small and subject to model details~\cite{Duerr:2011zd}, the black box theorem nonetheless implies that neutrinos are Majorana in nature. 
The signal for $0\nu\beta\beta$ decay is yet to be observed. Recent searches for $0\nu\beta\beta$ decay of $^{76}\mathrm{Ge}$ isotope constrains the half-life to $T_{1/2}(0\nu\beta\beta) \gtrsim 1.8 \times 10^{26}$ yr by GERDA~\cite{GERDA:2020xhi}, improving over previous measurements~\cite{Klapdor-Kleingrothaus:2000eir, EXO-200:2012pdt, Alenkov:2019jis}. Future experiments are expected to augment the sensitivity in lifetime by an order of magnitude or more~\cite{LEGEND:2021bnm}.

It is possible to extend the black box theorem to same-sign dileptons plus jets signals, $pp\to \ell^\pm\ell'^\pm+$~jets, at colliders. This process  violates lepton number by two units  and displays  a topology analogous to the $0\nu\beta\beta$ decay, as illustrated in Fig.~\ref{fig:black_box}. Hence, if we observe a signature of LNV with $\Delta L = 2$ at the LHC, this would imply Majorana nature for the neutrinos. 
While the $0\nu\beta\beta$ decay signals only probe effective  LNV interactions with two electrons $e^-e^-$, the LHC can  probe signals with electrons and muons in several combinations $(e^\pm e^\pm,\mu^\pm\mu^\pm,e^\pm\mu^\pm)$, increasing the sensitivity to a wider array of effective LNV interactions. The final state quarks could be of any flavor, CKM mixing would guarantee that neutrino mass would be induced by closing the quark loops.


\section{LNV signature in the type-II seesaw model}
\label{sec:type-II}

 The type-II seesaw model is a theoretical realization for non-zero neutrino masses that requires the existence of a $SU(2)_L$ triplet Higgs boson $\Delta$~\cite{Magg:1980ut, Schechter:1980gr, Cheng:1980qt, Mohapatra:1980yp} with hypercharge $Y = 1$. The relevant Lagrangian terms  leading to neutrino mass in this framework are 
 \begin{equation}
{\cal L} \supset     -Y_\nu l_L^T C i\sigma_2 \Delta l_L + \mu H^T i \sigma_2 \Delta^\dagger H+ \text{h.c.},
 \end{equation}
 where $Y_\nu$ is a $3\times 3$ complex symmetric matrix, $C$ is the charge conjugation operator, and $l_L^T=(\nu_L^T,e_L^T)$ stands for the three lepton doublets.  After electroweak symmetry breaking, the neutral component $\Delta^0$  acquires a vacuum expectation value  (VEV) $v_\Delta=\mu v^2_0/\sqrt{2}M_\Delta^2$, where  $M_\Delta$ is the mass of $\Delta^0$ and $(v_0^2+2 v_\Delta^2)\simeq (246~\text{GeV})^2$. Here $v_0$ is the VEV of the neutral component of the Higgs doublet. This symmetry-breaking pattern generates Majorana neutrino masses given by $m_\nu=\sqrt{2}Y_\nu v_\Delta$. One of the most distinguished phenomenological features of the type-II seesaw model is that the new triplet scalar fields can directly couple to the SM gauge bosons $(W^\pm,~Z,~\gamma)$, leading to exciting signatures at the LHC~\cite{Akeroyd:2007zv,Perez:2008zc,FileviezPerez:2008jbu,Melfo:2011nx}. The doubly charged member of the triplet scalar  can also contribute to $0\nu\beta\beta$ decay~\cite{Chakrabortty:2012mh}, however the corresponding amplitude is suppressed by a factor $Y_\nu v_{\Delta}/M_{\Delta}^{2} \sim m_\nu/M_{\Delta}^{2}$, compared to the amplitude for light neutrino exchange which goes as $m_\nu/\langle q^2\rangle$ with $\langle q^2 \rangle \sim (100~{\rm MeV})^2$,  leading to  null constraints  for $M_{\Delta} \sim \mathcal{O}(\mathrm{TeV})$~\cite{Chakrabortty:2012mh, Dev:2018sel}. 

  \begin{figure}[t!]
    \centering
    \includegraphics[width=.42\textwidth]{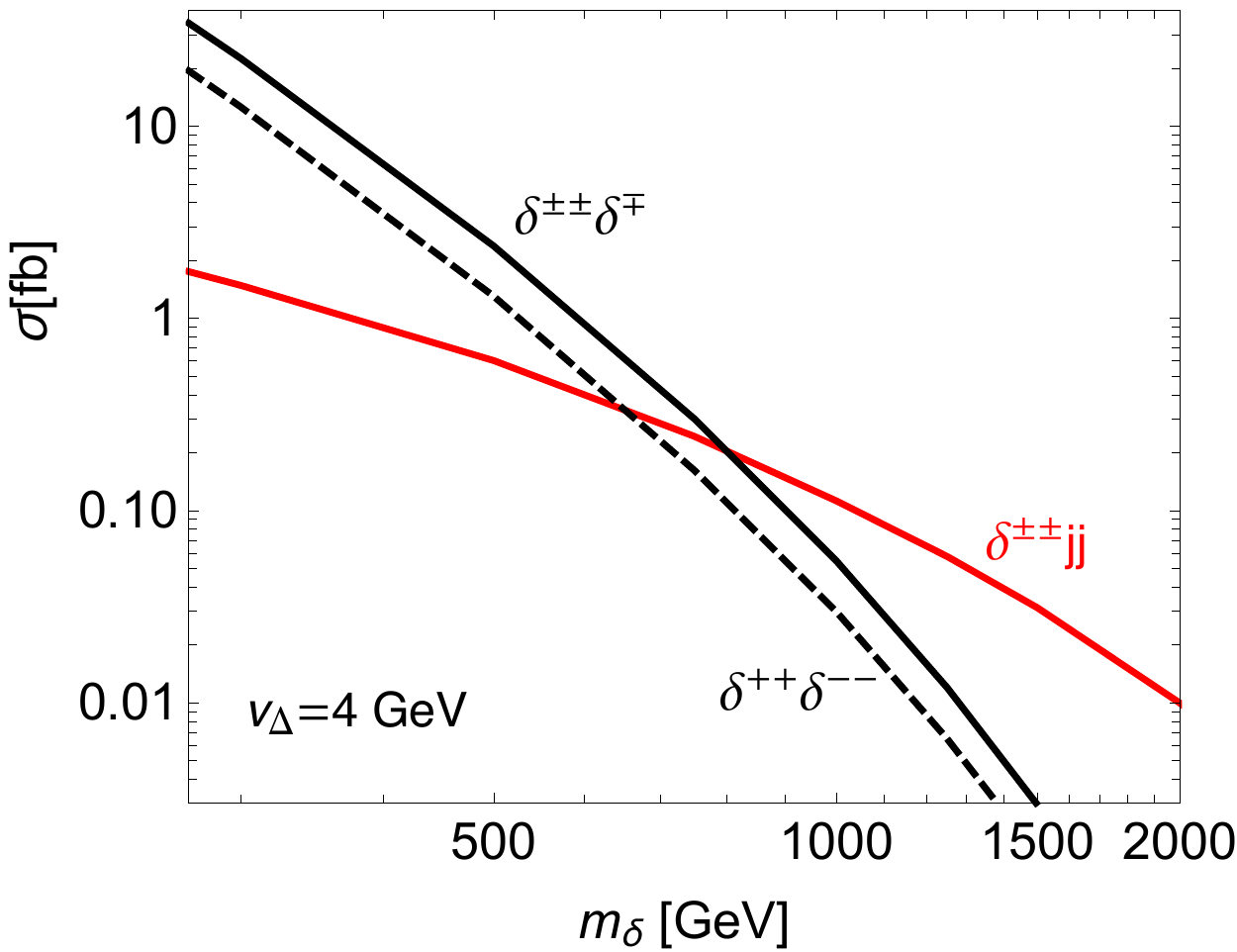}\hspace{1cm}
     \includegraphics[width=.4\textwidth]{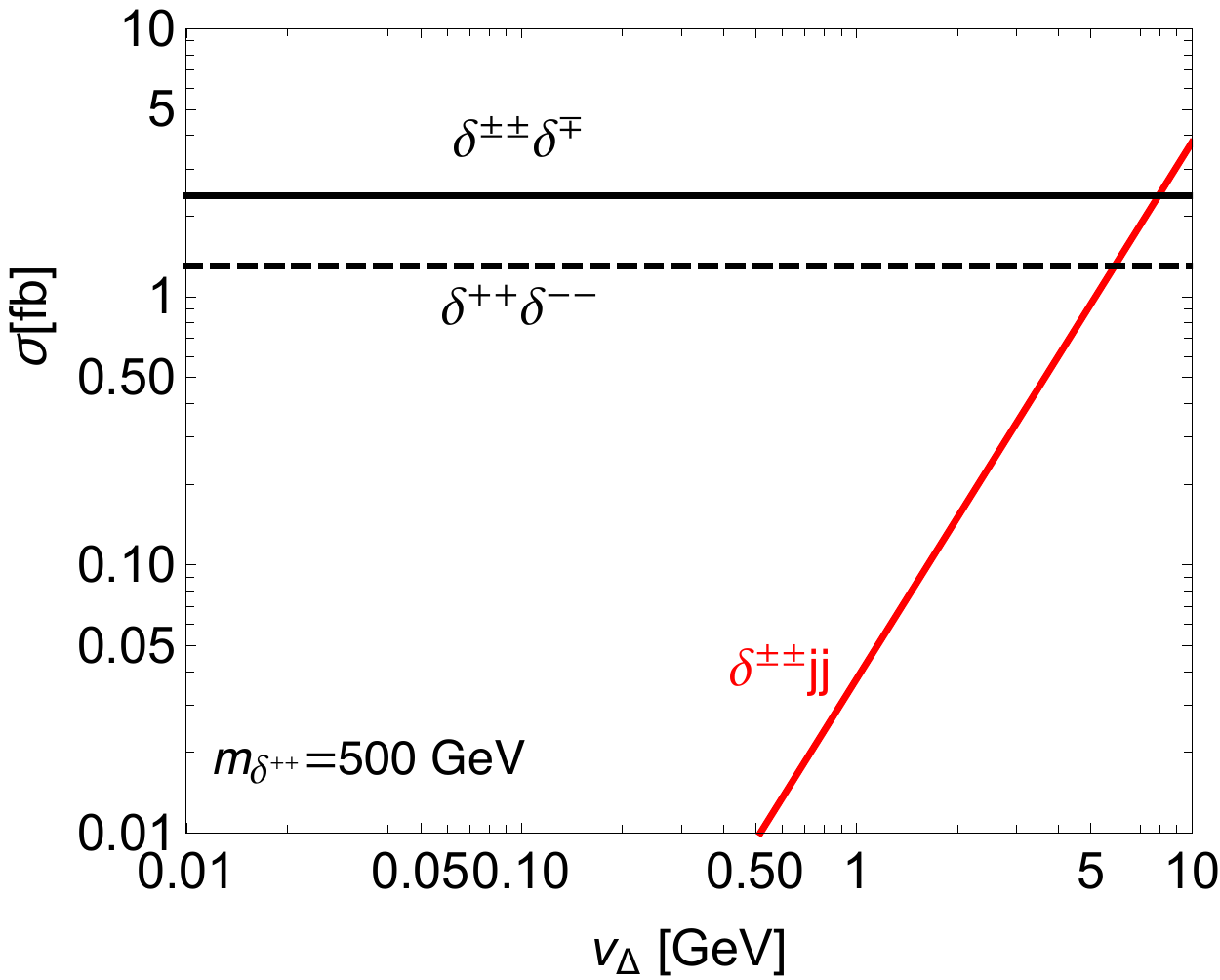}
    \caption{Cross-section for doubly charged Higgs boson pair production  $pp\rightarrow \delta^{\pm\pm}\delta^{\mp}$ (black solid), $pp\rightarrow \delta^{++}\delta^{--}$ (black dashed), and single production $pp\rightarrow \delta^{\pm\pm}jj$ (red)  as a function of $m_\delta$ (top panel) and $v_\Delta$ (bottom panel).  We assume the LHC at $\sqrt{s}=13$~TeV, $v_{\Delta}=4$~GeV on the top panel, and $m_\delta=500$~GeV on the bottom panel. Minimal selections $p_{Tj}>20$~GeV and $|\eta_j|<5$ are applied for the $\delta^{\pm\pm}jj$ process.}
    \label{fig:rate}
\end{figure}

\begin{figure*}[t!]
    \centering
    \includegraphics[width=0.20\textwidth]{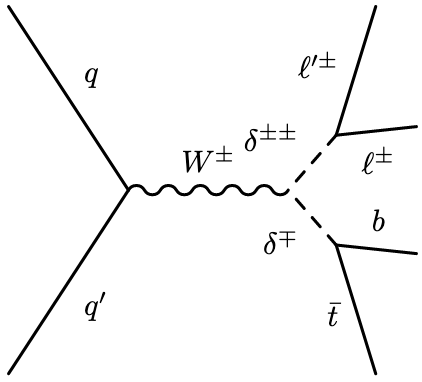}\hspace{0.5cm}\includegraphics[width=.20\textwidth]{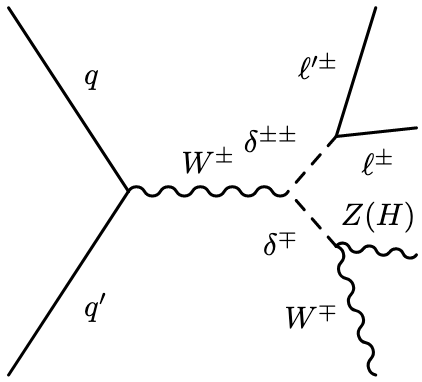} \vspace{0.2cm} \includegraphics[width=.18\textwidth]{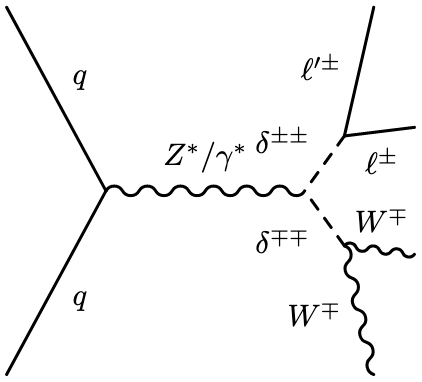}\hspace{0.5cm}\includegraphics[width=.18\textwidth]{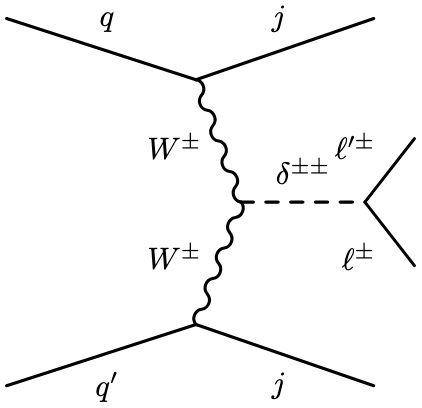}
    \caption{Representative set of Feynman diagrams contributing to the LNV signature $pp\rightarrow \ell^\pm\ell'^\pm$ + jets in the type-II seesaw model.}   
    \label{fig:feynamp_type_II}
\end{figure*}
 
 In this section, we study the LHC signals of LNV in the type-II seesaw scenario. In order to define LNV signals at the LHC, it is enough to observe final states with non-zero lepton number. In practice, this results in
the search for events with an excess of leptons or anti-leptons, ensuring that we do not lose track of the lepton number with neutrinos in the final state.  We will study a striking type-II seesaw signature of this type, characterized by the final state $\ell^\pm \ell'^\pm$ + jets.

 The leading production channels of doubly charged scalars at the LHC for the type-II scenario are via the $s$-channel electroweak process, displaying the dominant 
 associated production of double and single charged Higgs bosons $pp\rightarrow \delta^{\pm\pm}\delta^{\mp}$,  followed by  double charged Higgs pair production $pp\rightarrow \delta^{++}\delta^{--}$. In addition, the type-II seesaw model can also produce double charged scalars  via the vector boson fusion (VBF) mode, $pp\rightarrow   \delta^{\pm\pm}jj$.  In Fig.~\ref{fig:rate} (top panel), we illustrate the production cross-section  for these channels as a function of $m_\delta$. We observe that the VBF mode increases in relevance for large values of $m_\delta$, however, it still presents challenging event rates.    In Fig.~\ref{fig:rate} (bottom panel), we display the cross-section for the same channels as a function of the VEV of $\Delta$, $v_\Delta$.  While the Drell-Yan production channels  do not result in any $v_\Delta$ dependence,  the VBF  rate has a quadratic suppression, $\sigma(pp\rightarrow \delta^{\pm\pm}jj)\propto v_\Delta^2$. Electroweak precision observables, $\rho/T$-parameter, constraint the model parameters to the region with $v_\Delta\lesssim \mathcal{O}(1) $~GeV~\cite{Kanemura:2012rs}, limiting the phenomenological relevance of the VBF channel.

The LNV signatures $pp\rightarrow \ell^\pm\ell'^\pm$ + jets  arise from all three  leading production channels for doubly charged scalar(s) 
\begin{eqnarray}
pp &\rightarrow & \delta^{\pm\pm}\delta^\mp  \rightarrow \ell^\pm\ell'^\pm t b,~ \ell^\pm\ell'^\pm W^\mp Z/H\,, \label{eq:signal1}\\
pp &\rightarrow & \delta^{\pm\pm}\delta^{\mp\mp}  \rightarrow \ell^\pm\ell'^\pm W^\mp W^\mp\, ,  \label{eq:signal2}\\
pp &\rightarrow & \delta^{\pm\pm}jj  \rightarrow \ell^\pm\ell'^\pm jj\, ,
\label{eq:signal3}
\end{eqnarray}
where the associated  SM resonances $W$, $Z$, $t$, and $H$ decay hadronically. We display a representative set of Feynman diagrams for these processes in Fig.~\ref{fig:feynamp_type_II}. This signature displays uplifted event rates for values of $v_\Delta\sim 10^{-4}$~GeV. This parameter region benefits simultaneously from  singly and doubly charged scalars decays that are  proportional to the  neutrino Yukawa couplings $Y_\nu$ ($\delta^{\pm\pm} \rightarrow \ell^\pm\ell'^\pm$ and $\delta^\pm \rightarrow \ell^\pm\nu$) and  proportional to $v_\Delta$ ($\delta^{\pm\pm}\rightarrow W^\pm W^\pm$ and $\delta^\pm\rightarrow t b,~ W^\pm Z,~ W^\pm H$).

We perform the  Monte Carlo generation of the signal channels shown in  Eqs.~(\ref{eq:signal1}-\ref{eq:signal3}) with {\sc MadGraph5aMC@NLO} using the type-II {\sc FeynRules} model file~\cite{Alwall:2014hca,Fuks:2019clu}. Parton shower and hadronization effects are accounted for with {\sc Pythia8}~\cite{Sjostrand:2007gs}. Detector effects are simulated with {\sc Delphes3}~\cite{deFavereau:2013fsa}, using the default HL-LHC detector card~\cite{Cepeda:2019klc}. The same-sign dilepton searches suffer from large backgrounds from nonprompt leptons. Nonprompt leptons refer to leptons arising from decays of heavy flavor hadrons and jets misidentified as leptons. Since these background components are challenging to reliably simulate, we obtain the  background estimation from the same-sign dilepton plus jets search from  CMS to ensure a robust numerical estimation~\cite{Sirunyan:2020ztc}. This CMS study originally focuses on new physics interpretations in terms of supersymmetric models conserving or violating R-parity.

We start our analysis demanding exactly two same-sign leptons with transverse momenta $p_{T\ell}>25$~GeV and rapidity $|\eta_\ell|<2.5$ for electrons ($|\eta_\ell|<2.4$ for muons). We reject events with same (different) flavor leptons with mass $m_{\ell\ell}<12$~GeV (8~GeV), and $\slashed{E}_T<50$~GeV. Jets are defined with the anti-kt jet algorithm with radius $R=0.4$, $p_{Tj}>40$~GeV, and $|\eta_j|<2.4$. We require two or more jets in the event and  the scalar $p_{T}$ sum of all jets  $H_T>1125$~GeV. 

\begin{figure}[t!]
    \centering
    \includegraphics[width=.45\textwidth]{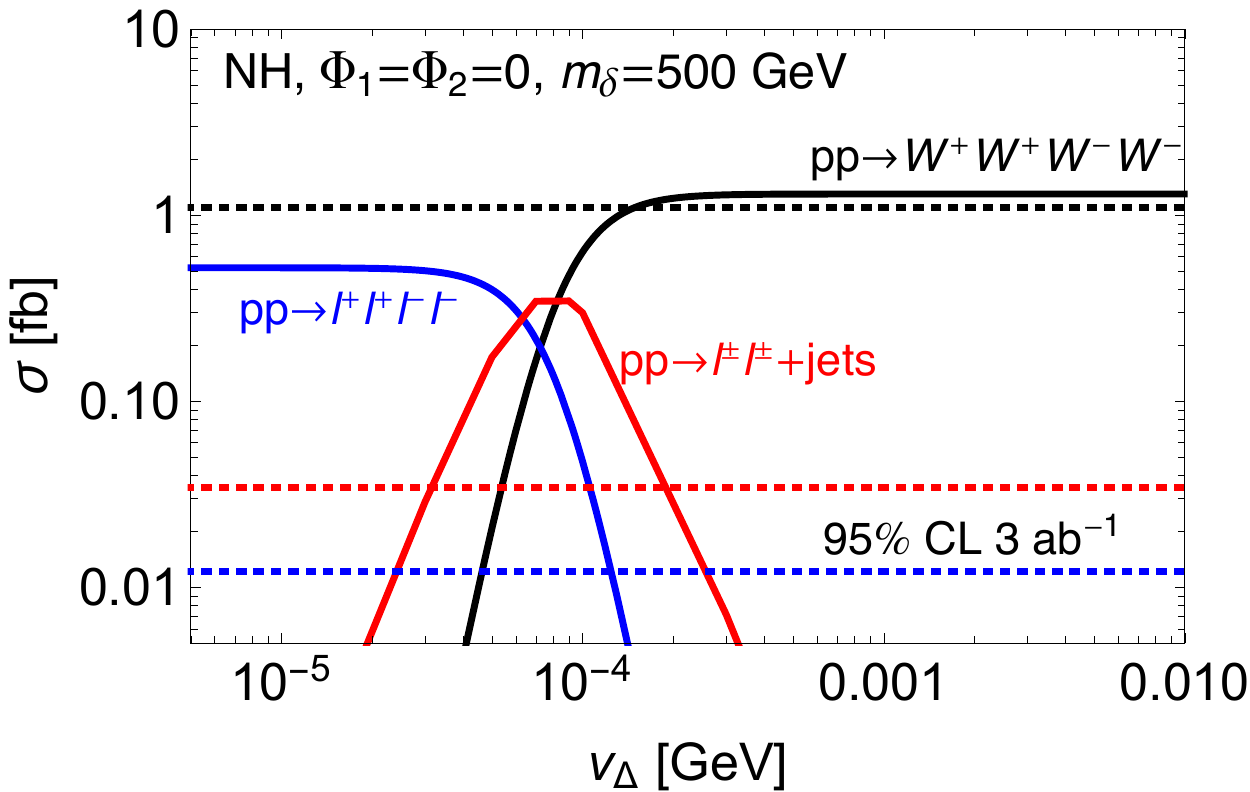}\hspace{1cm}
  \includegraphics[width=.45\textwidth]{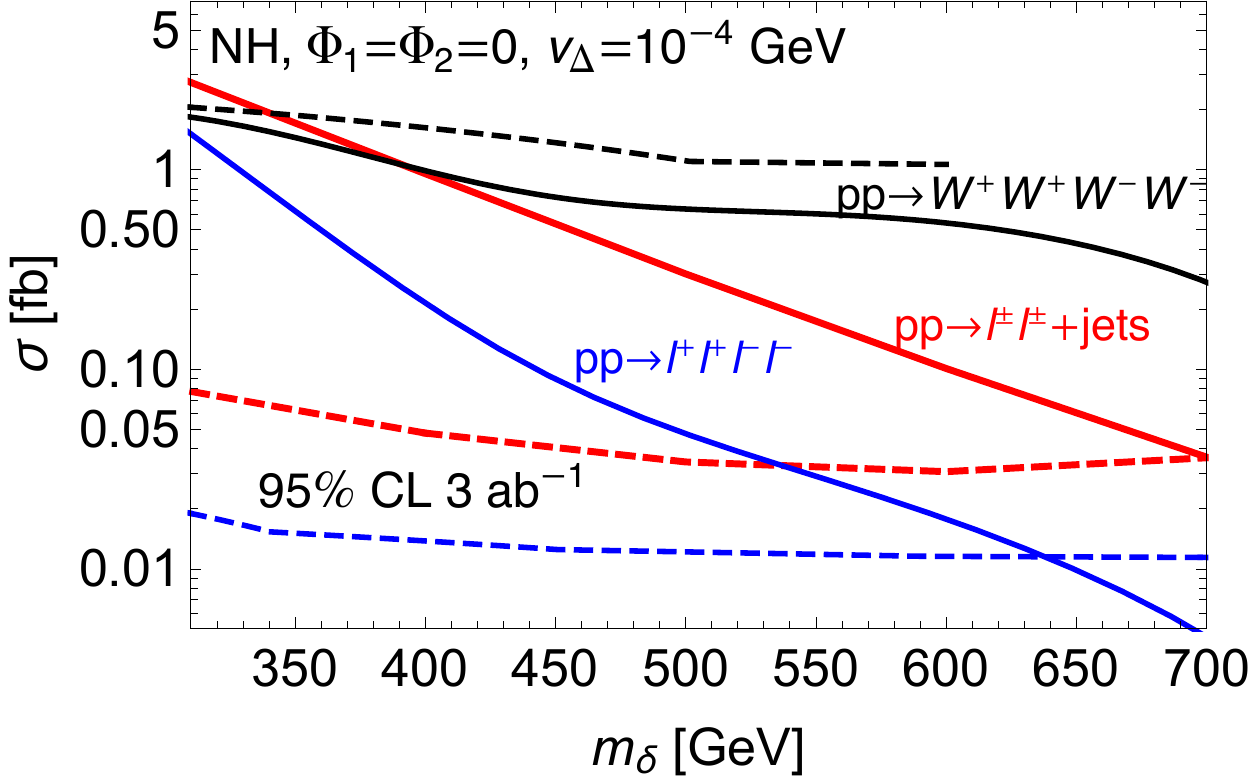}
    \caption{Cross-section for $pp\rightarrow \delta^{++} \delta^{--}\rightarrow \ell^+\ell^+\ell'^-\ell'^- $ with $\ell=e,\mu$ (blue solid), $pp\rightarrow \delta^{++} \delta^{--}\rightarrow W^+ W^+W^-W^- $ (black solid), and $pp\rightarrow \delta^{\pm\pm} \delta^{\mp\mp} + \delta^{\pm\pm} \delta^\mp +\delta^\pm jj\rightarrow \ell^\pm\ell^\pm+$~jets (red solid) in the Type II seesaw as a function of $v_\Delta$ (top panel) and $m_\delta$ (bottom panel). The 95\% confidence level limits for the high-luminosity LHC with  $3~\text{ab}^{-1}$ (dashed) are shown for all considered channels. 
    The results satisfy the neutrino global fit, in the normal mass hierarchy scenario (NH), presented in Ref.~\cite{Esteban:2018azc}. We assume the central values for the fit parameters along with $m_{\nu_1}=0.05$~eV, Majorana phases $\Phi_1=\Phi_2=0$, $m_{\delta^{++}}=m_{\delta^+}= m_\delta=500$~GeV (top panel), $v_\Delta=10^{-4}$~GeV  (bottom panel), and the LHC at $\sqrt{s}=13$~TeV.}
    \label{fig:limits-type2}
\end{figure}

In Fig.~\ref{fig:limits-type2}, we present the LHC  prediction for $pp\rightarrow \ell^\pm\ell'^\pm$ + jets (red-solid) in the normal mass hierarchy  scenario.  The results satisfy the neutrino global fit presented in Ref.~\cite{Esteban:2018azc}. We have taken the central values of the global fits presented in Ref.~\cite{Esteban:2018azc} in our analysis, along with $m_{\nu_1} = 0.05$ eV, and vanishing Majorana phases. We also assume that the triplet scalars have a common mass and  denote it as $m_\delta$. 
The CMS limit on the cross-section was scaled to the HL-LHC integrated luminosity $\mathcal{L}=3~\text{ab}^{-1}$ (red-dashed line)~\cite{Sirunyan:2020ztc}. We observe that our type-II signal displays relevant rates for moderate $v_\Delta$, excluding the parameter region  $3\times 10^{-5}~\text{GeV}\lesssim v_\Delta\lesssim 2\times 10^{-4}$~GeV at  95\% CL  for $m_\delta=500$~GeV. We can probe this  signal of LNV at the HL-LHC up to  $m_\delta=700$~GeV, corresponding to $v_\Delta=10^{-4}$~GeV. The dominant contributions for the signal are given by the channels in  Eqs.~(\ref{eq:signal1}) and~(\ref{eq:signal2}). The VBF channel presented in Eq.~(\ref{eq:signal3}) displays  subleading effects, being suppressed by $v_\Delta$, as shown in Fig.~\ref{fig:rate} (bottom panel). 

To illustrate the relevance of our LNV $\ell^\pm\ell'^\pm$ + jets search for the type-II seesaw, we present the ATLAS analyses on $\ell^+\ell^+\ell^-\ell^-$ (blue-dashed) and $W^+W^+W^-W^-$ (black-dashed), rescaling the limits to the HL-LHC luminosity~\cite{Aaboud:2017qph,ATLAS:2021jol}. While a possible observation of an excess of events in one of these two standard channels represents a clear sign of new physics, it does not phenomenologically translate into evidence of LNV as both channels display null lepton number in the final state. The correspondent signal generation was performed with the same Monte Carlo setup as for our $\ell^\pm\ell'^\pm$ + jets study. The signal cross-sections for $pp\rightarrow \delta^{\pm\pm}\delta^{\mp\mp}\rightarrow \ell^+ \ell^+ \ell^- \ell^- $ and $pp\rightarrow \delta^{\pm\pm}\delta^{\mp\mp}\rightarrow W^+ W^+ W^- W^- $ are shown in blue-solid and black-solid lines, respectively. Whereas the $4\ell$ signature thrives in the small $v_\Delta$ region and the $4W$ benefits from large $v_\Delta$, the $\ell^\pm\ell'^\pm$ + jets can probe intermediate values of $v_\Delta$ being complementary to both channels. Hence, the $pp\rightarrow \ell^\pm\ell'^\pm$ + jets search, in addition to providing a striking LNV signal, results in a competitive signature for intermediate $v_\Delta$ regimes for the type-II seesaw model.

\section{LNV signature in the Zee Model}
\label{sec:zee-model}

In this section, we study the projected sensitivity of LNV signals at the HL-LHC within the framework of the Zee model of neutrino masses~\cite{ZEE1980389}. The Zee model is perhaps the simplest extension of the SM that can generate non-zero neutrino masses radiatively at the one-loop level. The new physics responsible for inducing neutrino mass can be at the TeV-scale, since neutrino masses suffer from a loop suppression as well as a chiral suppression proportional to the charged lepton masses.   

The model extends the SM by introducing a second $SU(2)_{L}$ Higgs doublet $\Phi_{2}$ and a charged scalar singlet $\eta^{+}$. The SM-like Higgs doublet is denoted as $\Phi_{1}$. The two Higgs doublets $\{\Phi_{1},\Phi_{2}\}$ can be redefined to a new basis $\{H_{1},H_{2}\}$, where the neutral component of $H_{1}$ solely acquires a VEV $v$, in the so-called Higgs basis. The charged scalar doublet $H_{2}^{+}$ and singlet $\eta^{+}$ are composed of physical mass eigenstates $\{h^{+},H^{+}\}$, with $h^{+} = \cos\varphi~H_{2}^{+} - \sin\varphi~\eta^{+}$ and $H^{+} = \sin\varphi~H_{2}^{+} + \cos\varphi~\eta^{+}$. The mixing angle $\varphi$ is given by $\sin 2\varphi = -\sqrt{2}v\mu/(m_{H^{+}}^{2} - m_{h^{+}}^{2})$. Here, $\mu$ is the coefficient for the scalar cubic coupling term in the Higgs potential
\begin{equation}
V \supset \mu H_{1}^{i} H_{2}^{j}\eta^{-}\epsilon_{ij}
\label{eq:zee_V_muterm} + h.c.
\end{equation}

Following the notations in Ref.~\cite{Babu:2019mfe}, the leptonic and quark Yukawa Lagrangian for the BSM Higgs fields in the Higgs basis can be expressed as
\begin{eqnarray}
    \label{eq:zee_lepton_yukawa}
    - \mathcal{L}_{Y} &\supset& f_{\alpha\beta}L_{\alpha}^{i}L_{\beta}^{j}\epsilon_{ij}\eta^{+} + Y_{\alpha\beta}\tilde{H}_{2}^{i}L_{\alpha}^{j} \ell_{\beta}^{c}\epsilon_{ij} \\ 
    &+& \tilde{Y_u}_{\alpha\beta} H_{2}^{i}Q_{\alpha}^{j} u_{\beta}^{c}\epsilon_{ij} + {Y_d}_{\alpha\beta}\tilde{H}_{2}^{i}Q_{\alpha}^{j} d_{\beta}^{c}\epsilon_{ij} +  h.c., \nonumber
\end{eqnarray}
where $\{i,j\}$ are $SU(2)_{L}$ indices, $\{\alpha,\beta\}$ are the generation indices, $f_{\alpha\beta} = -f_{\beta\alpha}$ is an antisymmetric matrix in flavor space, 
$L$ and $Q$ denotes the left-handed lepton and quark doublets while $\ell^{c}$, $u^{c}$ and $d^{c}$ are the left-handed antileptons, up-type and down-type antiquarks, respectively. The Yukawa coupling matrices for the second Higgs doublet ($Y, \tilde{Y}_{u}, Y_{d}$) are complex asymmetric matrices and $\tilde{H}_{2} \equiv i \tau_{2} H_{2}^{\star}$. The charged lepton mass matrix $M_{\ell} = \tilde{Y} v/\sqrt{2}$ can be chosen to be diagonal such that $M_{\ell} = \left(m_{e},m_{\mu}, m_{\tau}\right)$ after electroweak symmetry breaking, without any loss of generality. (Here $\tilde{Y}$ is the Yukawa coupling matrix of $H_1$ to the charged leptons.) Note that we have allowed the most general set of Yukawa couplings in Eq. (\ref{eq:zee_lepton_yukawa}), including couplings of quarks with $H_2$.

The leptonic Yukawa interactions terms in Eq.~\eqref{eq:zee_lepton_yukawa}, together with the cubic coupling terms in Eq.~\eqref{eq:zee_V_muterm} induce explicit lepton number violation with $|\Delta L| = 2$ selection rule. This leads to non-zero neutrino masses induced at the one-loop level, with the mass matrix expressed as $m_{\nu} = \kappa(fM_{\ell}Y + Y^{T}M_{\ell}f^{T})$, where $\kappa = (1/16\pi^{2}) \sin2\varphi~{\rm log}(m_{h^{+}}^{2}/m_{H^{+}}^{2})$ is the one-loop suppression factor. Consistency with neutrino oscillation data requires the product of $f$ and $Y$ to be small, which can be realized by adopting $\mathcal{O}(1)$ values for $Y$ and very small $f$ couplings, $f \ll 1$, or \textit{vice versa}. Since our goal is to study the LHC signals of lepton number violation, we consider the latter scenario, $f \sim \mathcal{O}(1)$ and $Y \ll 1$, as the LNV interactions exclusively arise from the $f$-dependent couplings of $\eta^{+}$ in conjunction with the cubic scalar coupling $\mu$.

\begin{figure}
    \centering
    \includegraphics[width=0.2\textwidth]{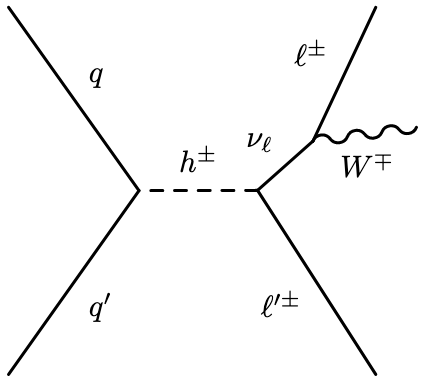}\includegraphics[width=0.2\textwidth]{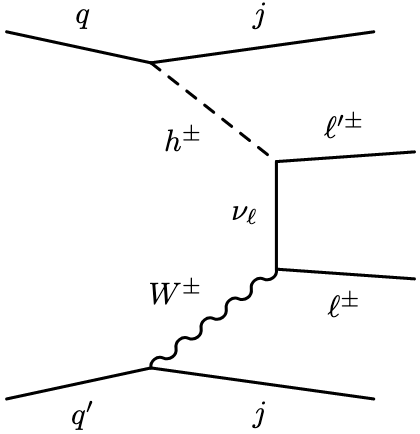}
    \caption{Feynman diagrams at the Born-level illustrating the leading $s$-channel (\textit{left panel}) and sub-leading $t$-channel (\textit{right panel}) production modes for singly charged Higgs $h^{\pm}$ in the Zee model, leading to LNV signatures $pp \to \ell^{\pm}\ell'^\pm$ + jets }
    \label{fig:feynamp_zee_model}
\end{figure}

As discussed previously in Sec.~\ref{sec:type-II}, the lepton number violating signature is characterized by two same-sign charged leptons plus jets, $pp \to \ell^{\pm}\ell^{\prime \pm}$ + jets. In the Zee model, such a final state can arise from LNV decays of the charged scalar $h^{+}$ on account of its $\eta^{+}$ admixture thanks to the cubic scalar coupling $\mu$ of Eq. (\ref{eq:zee_V_muterm}). Expanding the first term of the Lagrangian in Eq.~\eqref{eq:zee_lepton_yukawa}, and using Eq.~\eqref{eq:zee_V_muterm}, the LNV interaction terms for the $h^{+}$ can be expressed as follows:
\begin{eqnarray}
    -\mathcal{L}_{Y} \supset 2\sin\varphi ~h^{+}&[f_{e\mu}(\nu_{\mu}e& - \nu_{e}\mu)  + f_{e\tau}(\nu_{\tau}e - \nu_{e}\tau) \nonumber \\ + &f_{\mu\tau}(\nu_{\mu}\tau& - \nu_{\tau}\mu)] + h.c.
    \label{eq:hplus_lep_interactions}
\end{eqnarray}
We focus on the region of parameter space where $f_{e\mu} \sim \mathcal{O}(1) > f_{e\tau}, \,f_{\mu\tau}$, ignoring the implications from the latter two couplings for our collider study.  This is preferable since $f_{e\tau}$ and $f_{\mu\tau}$ would lead to  signatures with $\tau$ leptons in the final state. Both leptonic and hadronic decay modes of the $\tau$ result in neutrinos which can carry away an unknown lepton number, impeding the reconstruction of the total lepton number for the final state. On the other hand, $f_{e\mu}$ would allow LNV decays of $h^{\pm}$ into the same-sign electron-muon pair plus jets, $h^{\pm} \rightarrow e^{\pm}\mu^{\pm}(W^{\mp} \to jj)$, which can be fully reconstructed.  It should be noted that from the expression for the neutrino mass matrix $m_\nu$ given above, for any given value of $f_{e\mu}$ one can find a choice of other parameters of the model where neutrino oscillation data can be fitted \cite{Babu:2019mfe}. We also note that within our scenario ($Y\ll 1$), $f_{e\mu}$ is most stringently constrained by the lepton-hadron universality tests~\cite{ParticleDataGroup:2022pth,Nebot:2007bc}, which imposes the updated upper limit $|f_{e\mu}\sin\varphi|^{2} \leq 0.02 \left(m_{h^{+}}/\mathrm{TeV}\right)^{2}$. We shall impose this constraint in our analysis.

 \begin{figure}[t!]
    \centering
    \includegraphics[width=.45\textwidth]{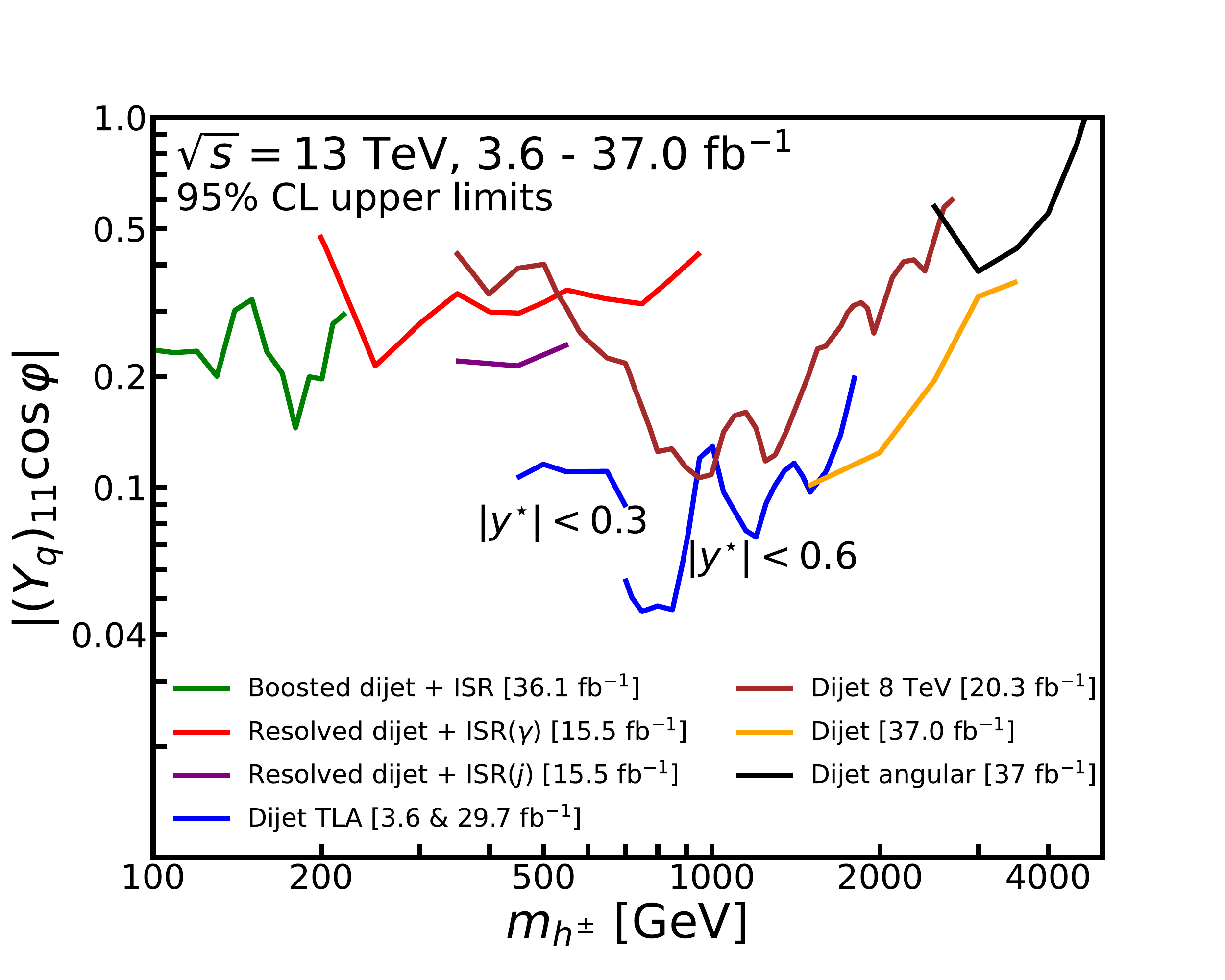}
    \caption{Upper limits at $95\%$ CL from dijet searches on $|(Y_{q})_{11}\cos\varphi|$ as a function of charged Higgs mass $m_{h^{\pm}}$ for the Zee model.}
    \label{fig:dijet_constraints}
\end{figure}

\begin{table}[!t]
    \centering
    \begin{tabular}{|c|c|c|c|c|} \hline 
     $m_{H}$ & \multirow{2}{*}{$|f_{e\mu}\sin\varphi|$} & \multirow{2}{*}{$|(Y_{q})_{11}\cos\varphi|$} &  $\Gamma_{h^{\pm}}$ & $\sigma(pp \to e^{+}\mu^{+}jj)$ \\ 
    $\rm{[GeV]}$ &  &  & [GeV] & [fb] \\ \hline 
    160 & 0.02 & 0.23  & 0.53  & 0.40 \\ 
    200 & 0.03 & 0.19  & 0.46 & 1.4 \\
    400 & 0.06 & 0.21  & 1.3 &  10.3 \\
    600 & 0.08 & 0.11  & 0.91 &  13.6 \\
    800 & 0.11 & 0.05  &  0.86 &  3.1 \\
    1030 & 0.15 & 0.11  & 1.7  &  7.2 \\
    1500 &  0.21 & 0.10 & 2.3 &  3.0 \\
    1900 & 0.27 & 0.12 & 3.5 & 2.1 \\
    3000 &  0.42 & 0.33  & 22 & 0.73 \\
    4000 & 0.57 & 0.55  & 77 & 0.16 \\
    \hline 
    \end{tabular}
    \caption{Highest allowed values for $|f_{e\mu}\sin\varphi|$ from lepton-hadron universality constraints~\cite{ParticleDataGroup:2022pth} and $|(Y_{q})_{11}\cos\varphi|$ from dijet searches at the LHC~\cite{PhysRevD.91.052007, ATLAS:2017eqx, ATLAS:2018hbc, ATLAS-CONF-2016-070, ATLAS:2019wdu, ATLAS:2018qto} for several charged Higgs masses $m_{h^{\pm}}$ in the Zee model, with $f_{e\mu} \sim \mathcal{O}(1) > f_{e\tau}, f_{\mu\tau}$, $Y_{\ell\ell} << 1$, $(Y_{q})_{11/33} \sim \mathcal{O}(1)$, and $(Y_{q})_{22},(Y_{q})_{ij}(i\neq j) =0$. We also restrict to $|f_{e\mu}\sin\varphi| < 0.1$ and $|(Y_{q})_{11}\cos\varphi| < 1$ in order to avoid perturbativity issues. The total decay width for the charged scalar is also shown along with the tree-level production rates for the LNV signal, $pp \to e^{\pm}\mu^{\pm}jj$, at $\sqrt{s}=13~$TeV.}
    \label{tab:zee_model_cs_val}
\end{table}

We next focus on the production of $h^{\pm}$ at the LHC. By virtue of its mixing with the charged doublet Higgs $H_{2}^{+}$ which has Yukawa couplings to quarks, $h^{+}$ can be resonantly produced in the quark fusion channel.
The $h^\pm$ field would decay predominantly into   
$\ell^\pm \nu$, which is however not useful for $L$-violation studies owing to the presence of neutrinos.  However, $h^\pm$ has a subleading decay into  $e^{\pm}\mu^{\pm}W^{\mp}$, which gives significant event rates. For example, at $m_{h^{\pm}} = 1~$TeV, the branching ratios for the $h^{\pm} \to \ell^{\pm}\nu$ and $e^{\pm}\mu^{\pm}W^{\mp}$ decay modes are roughly $24\%$ and $1.2\%$, respectively, considering $|f_{e\mu}\sin\varphi| \sim 0.1$ and $|(Y_{u/d})_{11}\cos\varphi| = |(Y_{u/d})_{33}\cos\varphi| \sim 0.1$.  Observable LNV signature can thus arise, even with the small branching ratio of order 1\%, in the resonant $h^{\pm}$ production channel $pp \to h^{\pm} \to e^{\pm}\mu^{\pm}W^{\mp}$ with the $W^{\mp}$ boson decaying hadronically. 
Sub-leading contributions emerge from $h^{\pm}$ mediation in the $t$-channel. Representative Feynman diagrams at the Born-level for $s$- and $t$-channel $h^{\pm}$ exchange are shown in Fig.~\ref{fig:feynamp_zee_model}. It is worth pointing out that the topology of the right panel diagram of Fig.~\ref{fig:feynamp_zee_model} is the so-called vector-scalar exchange contribution for neutrinoless double beta decay \cite{Babu:1995vh}, if the final state leptons are two electrons. 
We consider both production modes in the present analysis. In order to avoid the stringent flavor constraints from meson decays~\cite{FlaviaNetWorkingGrouponKaonDecays:2008hpm, Mahmoudi:2016mgr, HFLAV:2022pwe}, all entries for $Y_{u/d}$ are set to zero, except for first $(Y_{u/d})_{11}$)and third $(Y_{u/d})_{33}$ generation diagonal entries.\footnote{The diagonal entries for the third-generation Yukawa couplings are considered non-zero for the sake of generality, set to $|(Y_{q})_{33}\cos\varphi| = 0.1$ in our numerical analysis.} Thus, in the present scenario, the charged Higgs $h^{\pm}$ contributes to leptonic decay, credits to  $f_{e\mu} = -f_{\mu e}$, and hadronic decay into first and third-generation quarks. Typically, this implies constraints from pion decay and nuclear $\beta$ decay. However, $h^{\pm}$ couples with charged and neutral leptons from different flavors due to the antisymmetric structure of $f$, leading to no interference with the $W$ boson-mediated nuclear $\beta$ decay. The constraints from beta decay are therefore very weak for $h^\pm$ masses of order TeV. Furthermore, pseudoscalar interactions are necessary to induce contributions to charged pion decay, which would be highly constraining. A pseudoscalar coupling of $h^\pm$ of the form $y_p \cos\phi \,(\overline{u} \gamma_5 d) \,h^+$ would lead a constraint of $|y_p \,f_{e\mu} \cos\phi \sin\phi| \leq 5 \times 10^{-4}\, (m_{h^\pm}/{\rm TeV})^2$ from $\Gamma(\pi \rightarrow e \nu_\mu)/\Gamma(\pi \rightarrow \mu \nu_\mu)$ measurement \cite{Campbell:2003ir}. This constraint can be evaded by taking $(Y_{d})_{11} = (\tilde{Y}_{u})_{11} \equiv (Y_{q})_{11}$, in which case the interaction of $h^\pm$ with the quarks are purely scalar with $y_p = 0$.  Therefore, constraints from pion decay and $\beta$ decay can be safely ignored. Under the previously discussed assumptions, the only constraints on the $u\bar{d}h^{+}$ couplings ($\propto |(Y_{q})_{11}\cos\varphi|$) that govern $h^{\pm}$ production at the LHC arise from resonant dijet searches at the LHC. 

We translate the current exclusion limits from several relevant dijet searches onto the $\{m_{h^{+}}, (Y_{q})_{11}\cos\varphi\}$ plane for the Zee model. In Fig.~\ref{fig:dijet_constraints}, we show the limits from dijet searches performed using LHC $\sqrt{s}=8~$TeV data at $\mathcal{L}=20.3~\rm{fb}^{-1}$~\cite{PhysRevD.91.052007} (brown) and $\sqrt{s}=13~$TeV data at $\mathcal{L}=37~\rm{fb}^{-1}$~\cite{ATLAS:2017eqx} (orange). The green and red (purple) contours show the limits from boosted dijet $+$ ISR searches~\cite{ATLAS:2018hbc} and resolved dijet $+$ $\gamma~(j)$-ISR searches~\cite{ATLAS-CONF-2016-070}. The aforesaid search analyses consider a sliding window-fit of the dijet invariant mass distribution to estimate the background, thus, making them sensitive below a certain width-to-mass ratio $\Gamma_{h^{+}}/m_{h^{+}}$. The dijet limits are valid up to $\Gamma_{h^{+}}/m_{h^{+}} \leq 15\%$, while the dijet $+$ ISR search limits are valid only up to $\Gamma_{h^{+}}/m_{h^{+}} \leq 10\%$~\cite{ATLAS:2019wdu}. We also show the limits from trigger-object-level (TLA) dijet~\cite{ATLAS:2018qto}~(blue) and dijet angular analysis~\cite{ATLAS:2017eqx}~(black). The TLA dijet analysis focuses on two distinct selection criteria, $|y^{*}| < 0.3$ and $|y^{*}| < 0.6$, where $y^{*} = (y_{1} - y_{2})/2$ with $y_{1}$ and $y_{2}$ being the pseudorapidity of the highest and second-highest $p_{T}$ jets at the trigger-level. The TLA dijet analysis with $|y^{*}| < 0.3$ is sensitive up to $\Gamma_{h^{+}}/m_{h^{+}} \leq 10\%$, while the TLA dijet analysis assuming $|y^{\star}| < 0.6$ is sensitive only up to $\Gamma_{h^{+}}/m_{h^{+}} \leq 7\%$~\cite{ATLAS:2019wdu}. The angular dijet analysis is sensitive for wider resonances and is valid up to $\Gamma_{h^{+}}/m_{h^{+}} \leq 50\%$. 

 \begin{figure}[!t]
    \centering
    \includegraphics[width=.45\textwidth]{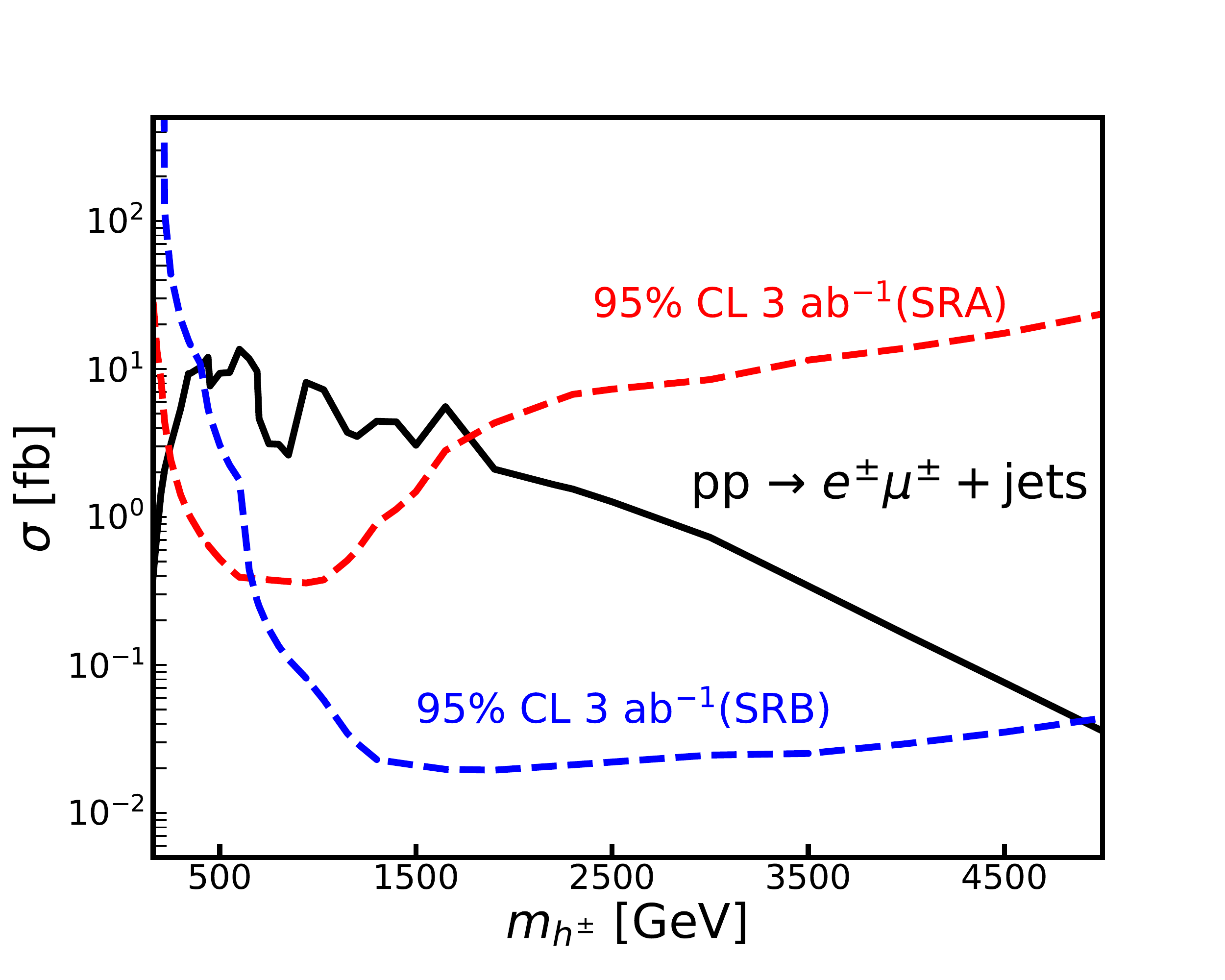}
    \caption{Cross-section for same-sign dilepton plus jets production $pp\to e^\pm \mu^\pm$ + jets (black-solid) as a function of the charged Higgs mass $m_{h^\pm}$ at the $\sqrt{s}=13$~TeV LHC. The 95\% confidence level limits for the high-luminosity LHC with $3$~ab$^{-1}$ (dashed) are shown for signal regions SRA [$H_{T} \subset [300,1125]~$GeV] (red) and SRB [$H_{T} > 1125~$GeV] (blue).}
    \label{fig:zee_rate}
\end{figure}

In Table~\ref{tab:zee_model_cs_val}, we show the highest allowed values for $|f_{e\mu}\sin\varphi|$ and $|(Y_{q})_{11}\cos\varphi|$ from lepton-hadron universality constraints and dijet bounds, respectively, for various charged Higgs masses. We note that the respective couplings have been restricted to $|f_{e\mu}\sin\varphi| < 0.1$ and $|(Y_{q})_{11}\cos\varphi| < 1$ in order to avoid endangering perturbativity, thus making our results conservative estimates. The total decay width for $h^{\pm}$ and the truth-level leading order cross-section for the LNV signal $pp \to e^{\pm}\mu^{\pm}jj$ at $\sqrt{s}=13~$TeV, computed using {\sc MadGraph5aMC@NLO}~\cite{Alwall:2014hca}, are also shown. 

\begin{figure*}[!t]
    \centering
    \includegraphics[width=0.32\textwidth]{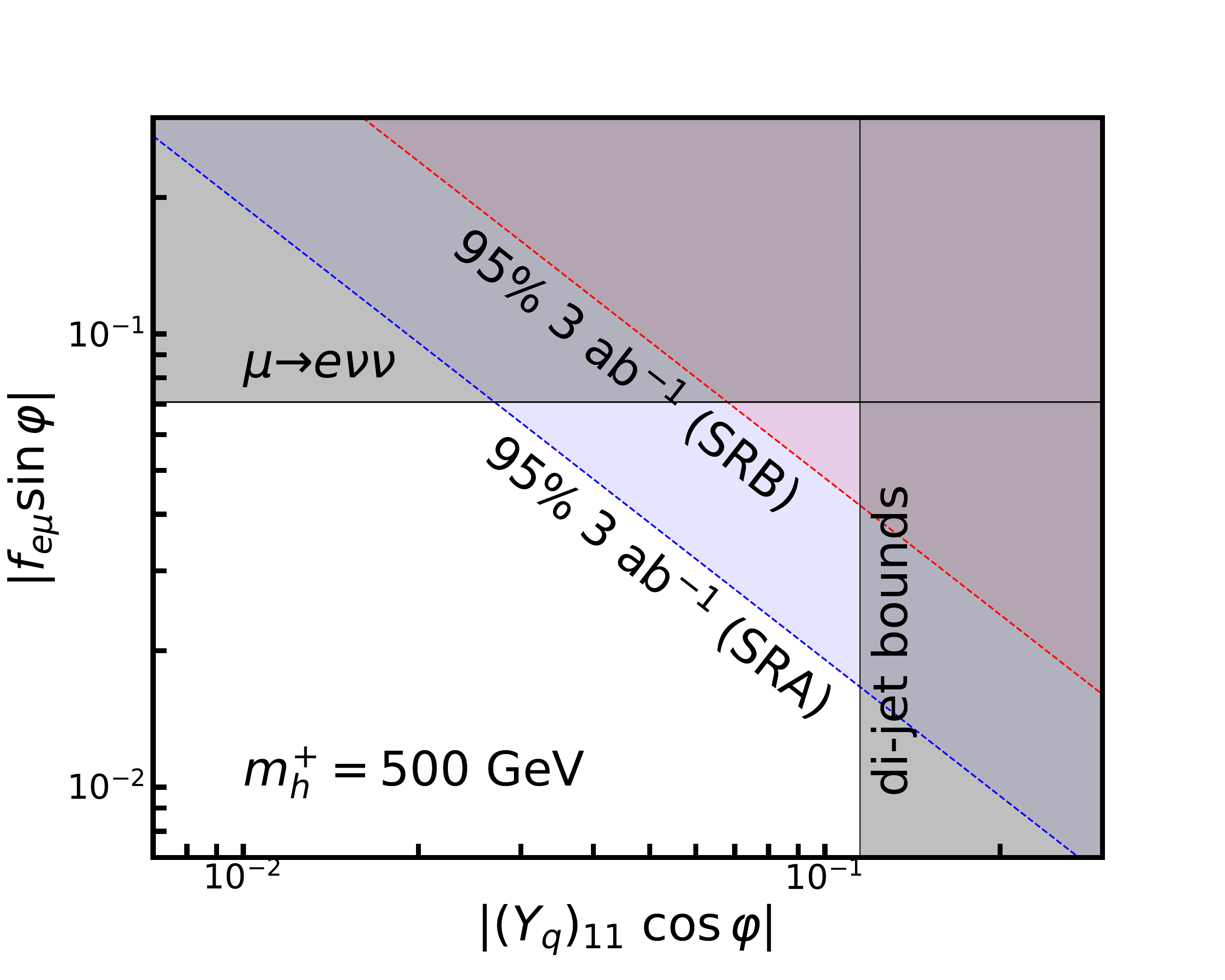}\includegraphics[width=0.32\textwidth]{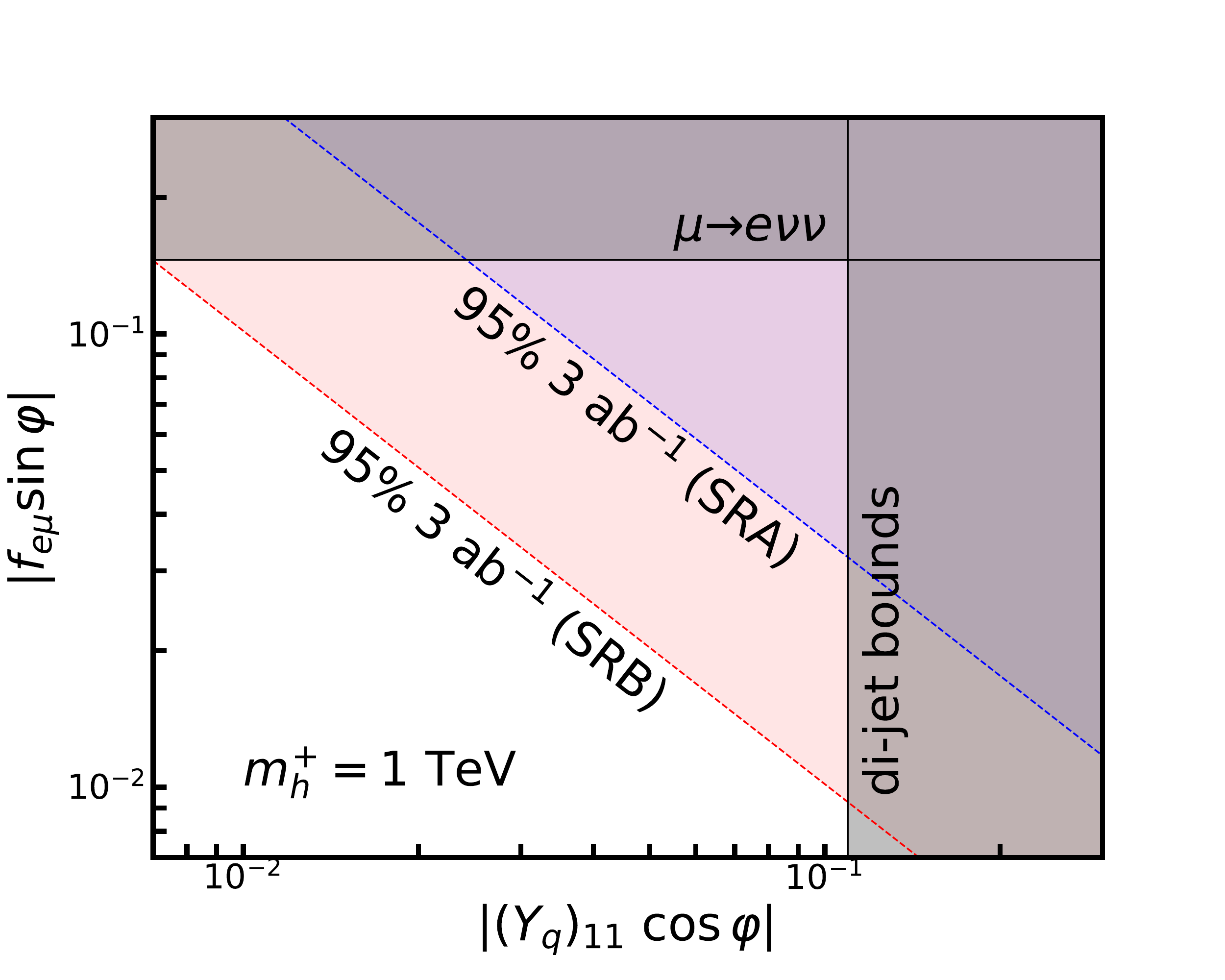}\includegraphics[width=0.32\textwidth]{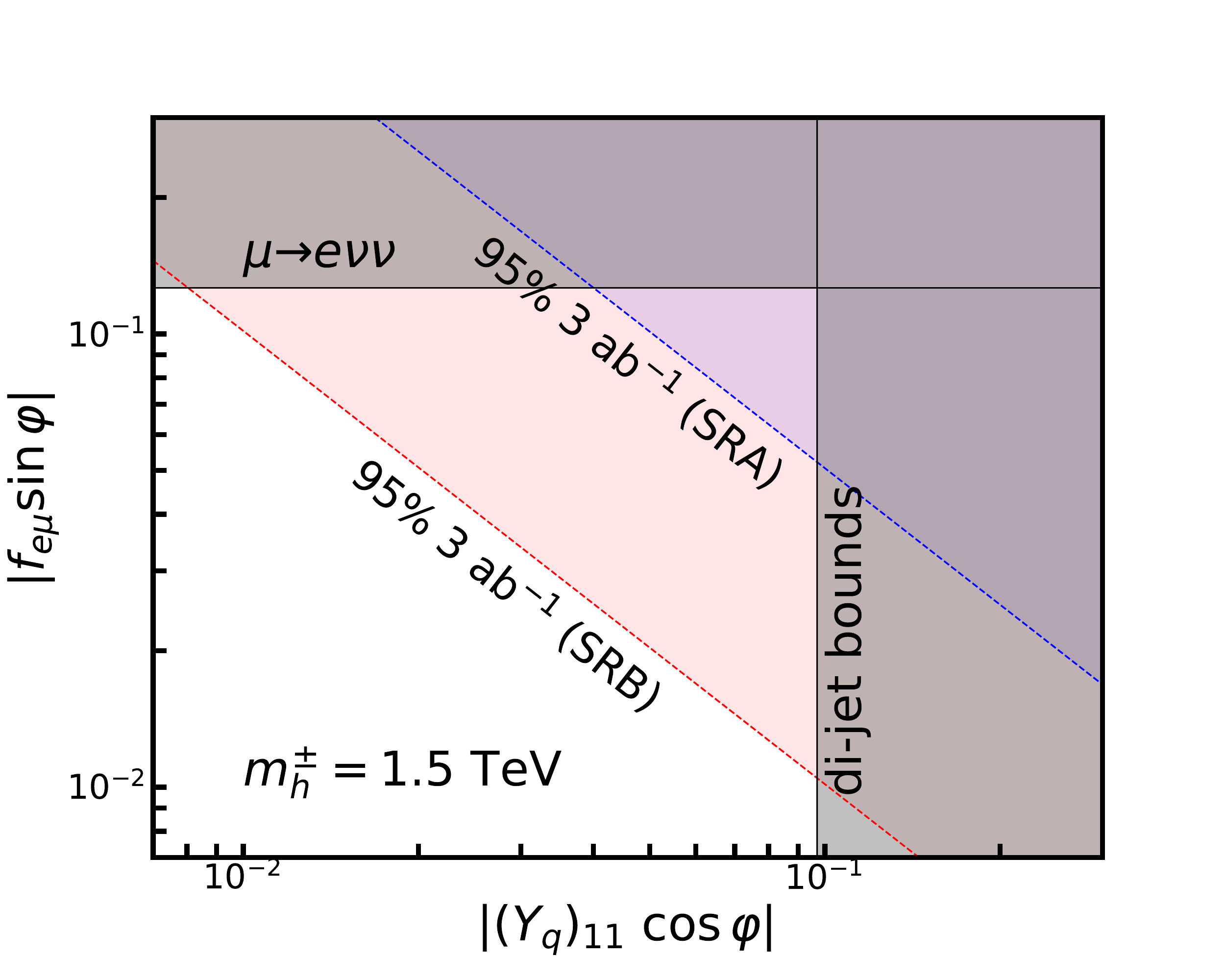}
    \caption{Projected exclusion regions in the plane of $|f_{e\mu}\sin\varphi|$ and $|(Y_{q})_{11}\cos\varphi|$ in the Zee model from searches in the lepton number violating $pp \to e^{\pm}\mu^{\pm}+$jets final state in signal region SRA (blue-shaded) and SRB (red-shaded) at the HL-LHC at $95\%$ CL. Three different charged Higgs masses are considered, $m_{h^{+\pm}}= (0.5,~1,~1.5)$ TeV. Current upper limits on $|f_{e\mu}\sin\varphi|$ from $\mu \to e \nu\nu$ decay, $|f_{e\mu}\sin\varphi|^{2} \leq 0.02 \times \left(m_{h^{\pm}}/\mathrm{TeV}\right)^{2}$~\cite{ParticleDataGroup:2022pth}, and on $|(Y_{q})_{11}\cos\varphi|$ from dijet searches at the LHC (see Fig.~\ref{fig:zee_rate}) are shown (grey-shaded).}
    \label{fig:zee_hllhc_exclusion}
\end{figure*}

The $pp \to e^{\pm}\mu^{\pm}$ + jets signal events in the Zee model are simulated using the analysis setup described in Sec.~\ref{sec:type-II}. Similar event selection criteria are also adopted, except that we consider two signal regions, one with $H_{T} \subset [300,1125]~$GeV and another with $H_{T} > 1125~$GeV, motivated by the CMS analysis for same-sign leptons plus jets~\cite{Sirunyan:2020ztc}. We refer to the two signal regions as SRA and SRB, respectively. It is worth mentioning again that the requirement of low missing energy ($\slashed{E}_{T}<50~$GeV) is crucial to avoid neutrinos which could carry away the lepton number, but in practice, many of our signal events fail this cut with our simulated detector resolution.

In Fig.~\ref{fig:zee_rate}, we present the cross-sections for the LNV signal $pp \to e^{\pm}\mu^{\pm}$ + jets (black-solid) at the $\sqrt{s}=13~\rm{TeV}$ LHC. We reiterate that the leading contribution to the LNV signal arises from resonant $h^{\pm}$ production in the $s$-channel. Note that the resonant charged Higgs production cross-section is found to be roughly consistent with that given in Ref.~\cite{Egana-Ugrinovic:2021uew} for the case of two Higgs doublet model.  The projected sensitivities at the HL-LHC in SRA (red-dashed) and SRB (blue-dashed), is computed by extrapolating the CMS limits~\cite{Sirunyan:2020ztc} for the respective signal regions through luminosity scaling. Signal regions SRA and SRB play a complementarity role, with the former being optimal at lower charged Higgs mass, $m_{h^{\pm}} \lesssim 800~$GeV. 

We observe that the LNV signature $pp \to e^{\pm}\mu^{\pm}$ + jets in the Zee model can be probed at the HL-LHC in signal region SRA for $m_{h^\pm} \subset [300,1550]~$GeV, and in SRB for $m_{h^\pm} \subset [500,4800]~$GeV, at $95\%$ CL. It is important to note that the LNV signal features a relatively small missing transverse energy $\slashed{E}_{T}$ irrespective of $m_{h^{\pm}}$, resulting from hadronization of the $b$ and light flavored quarks. As discussed previously, smaller $\slashed{E}_{T}$ values are a characteristic feature of the LNV signature, since the same-sign dilepton plus jets signal will be typically accompanied by neutrinos in the absence of lepton number violating couplings leading to relatively higher $\slashed{E}_{T}$ distributions. It is also worth noting that $h^{\pm}$ can be probed at the LHC in DY production, $pp \to \gamma^{*}/Z \to h^{\pm}h^{\mp} \to \ell^{\pm}\ell^{\prime\mp}+\slashed{E}_{T}$. However, this channel is weakly constrained at the LHC~\cite{Babu:2019mfe} by stau ($\tilde{\tau}$) pair production searches~\cite{CMS:2018yan}. 

We cast the results into the plane of $|f_{e\mu}\sin\varphi|$ and $|(Y_{q})_{11}\cos\varphi|$ in Fig.~\ref{fig:zee_hllhc_exclusion} to evaluate the projected reach at the HL-LHC for three different charged Higgs masses, $m_{h^{\pm}} = 0.5, 1, 1.5~$TeV. The grey-shaded region represents the parameter space excluded by current constraints. The blue and red-shaded areas fall within the projected reach of LNV searches in the $pp \to e^{\pm}\mu^{\pm}$ + jets channel at the HL-LHC at $95\%$ CL for signal regions SRA and SRB, respectively. We observe that the HL-LHC will be able to extend the reach further beyond the current sensitivity. Besides, as discussed previously, at lower charged Higgs masses, $m_{h^{\pm}} \sim 500~\mathrm{GeV}$, SRA is comparatively more sensitive than the SRB. Their relative impact is reversed in the case of $m_{h^{\pm}} \sim 1$ and $1.5~\mathrm{TeV}$.


\section{Summary}
\label{sec:summary}

Whether lepton number is a broken symmetry of nature or not is one of the profound questions in the SM. It strongly correlates with the question regarding the nature of neutrinos, credits to the black box theorem. Various SM  extensions have attempted to address these issues. The type-II seesaw model and the Zee model are compelling minimal extensions that generate small Majorana masses for the neutrinos implying LNV by two units. In a favorable range of parameters, the new scalar particles predicted by these models can be accessed at the LHC while being consistent with neutrino oscillation data. 

In this paper we  analyzed the HL-LHC prospects for lepton number violation signature, $pp \to \ell^{\pm}\ell'^{\pm}$ + jets, arising in the type-II seesaw model and the Zee model through detailed collider studies. In the type-II seesaw model, we considered three leading production channels for $\delta^{\pm\pm}$ that result  in final states with $L$-violation, namely Drell-Yan production of double and single-charged scalars, $pp \to \delta^{\pm\pm}\delta^{\mp}$, pair production of double charged scalars $pp \to \delta^{\pm\pm}$, and VBF production $pp \to \delta^{\pm\pm}jj$. We performed a collider study using the signal regions from the CMS analysis~\cite{Sirunyan:2020ztc} for the same-sign dilepton plus jets. The background estimates at the HL-LHC are extrapolated from the CMS measurements. Searches in the $pp \to \ell^{\pm}\ell'^{\pm}$ + jets results in a compelling signature for lepton number violation at the HL-LHC at $95\%$ CL, at intermediate $v_{\Delta} \sim 10^{-4}$ GeV. The search potential of the LNV signal is also shown to be complementary to the sensitivity of the standard $\delta^{\pm\pm}$ searches in the $4\ell$ and $4W$ final states. 

An analogous search strategy is adopted to analyze the HL-LHC sensitivity for the lepton number violating signal, $pp \to e^{\pm}\mu^{\pm}$ + jets, in the Zee model of neutrino mass. The LNV signal originates from decays of $h^{\pm}$, with the leading production process being resonant $h^{\pm}$ production via quark fusion. Here again, we observe that the HL-LHC will be able to probe the LNV signal at $95\%$~CL through searches in the $pp \to e^{\pm}\mu^{\pm}$ + jets channel, for a wide range of charged Higgs masses, $m_{h^{\pm}} \subset [0.3, 4.8]~$TeV. Notably, our results clearly demonstrate that the $L$-violating signature, $pp \to \ell^{\pm}\ell^{\prime\pm}$ + jets, in the type-II seesaw and the Zee Model, has the potential to be discovered at the HL-LHC. 


\section*{Acknowledgements}
This work is supported by the U.S. Department of Energy  under grant number DE-SC0016013. Some computing for this project was performed at the High Performance Computing Center at Oklahoma State University, supported in part through the National Science Foundation grant OAC-1531128.

\bibliography{mybib}

\begin{thebibliography}{113}
\expandafter\ifx\csname natexlab\endcsname\relax\def\natexlab#1{#1}\fi
\expandafter\ifx\csname bibnamefont\endcsname\relax
  \def\bibnamefont#1{#1}\fi
\expandafter\ifx\csname bibfnamefont\endcsname\relax
  \def\bibfnamefont#1{#1}\fi
\expandafter\ifx\csname citenamefont\endcsname\relax
  \def\citenamefont#1{#1}\fi
\expandafter\ifx\csname url\endcsname\relax
  \def\url#1{\texttt{#1}}\fi
\expandafter\ifx\csname urlprefix\endcsname\relax\def\urlprefix{URL }\fi
\providecommand{\bibinfo}[2]{#2}
\providecommand{\eprint}[2][]{\url{#2}}

\bibitem[{\citenamefont{Weinberg}(1979)}]{Weinberg:1979sa}
\bibinfo{author}{\bibfnamefont{S.}~\bibnamefont{Weinberg}},
  \bibinfo{journal}{Phys. Rev. Lett.} \textbf{\bibinfo{volume}{43}},
  \bibinfo{pages}{1566} (\bibinfo{year}{1979}).

\bibitem[{\citenamefont{Minkowski}(1977)}]{Minkowski:1977sc}
\bibinfo{author}{\bibfnamefont{P.}~\bibnamefont{Minkowski}},
  \bibinfo{journal}{Phys. Lett. B} \textbf{\bibinfo{volume}{67}},
  \bibinfo{pages}{421} (\bibinfo{year}{1977}).

\bibitem[{\citenamefont{Gell-Mann et~al.}(1979)\citenamefont{Gell-Mann, Ramond,
  and Slansky}}]{GellMann:1980vs}
\bibinfo{author}{\bibfnamefont{M.}~\bibnamefont{Gell-Mann}},
  \bibinfo{author}{\bibfnamefont{P.}~\bibnamefont{Ramond}}, \bibnamefont{and}
  \bibinfo{author}{\bibfnamefont{R.}~\bibnamefont{Slansky}},
  \bibinfo{journal}{Conf. Proc. C} \textbf{\bibinfo{volume}{790927}},
  \bibinfo{pages}{315} (\bibinfo{year}{1979}), \eprint{1306.4669}.

\bibitem[{\citenamefont{Glashow}(1980)}]{Glashow:1979nm}
\bibinfo{author}{\bibfnamefont{S.}~\bibnamefont{Glashow}},
  \bibinfo{journal}{NATO Sci. Ser. B} \textbf{\bibinfo{volume}{61}},
  \bibinfo{pages}{687} (\bibinfo{year}{1980}).

\bibitem[{\citenamefont{Yanagida}(1979)}]{Yanagida:1979as}
\bibinfo{author}{\bibfnamefont{T.}~\bibnamefont{Yanagida}},
  \bibinfo{journal}{Conf. Proc. C} \textbf{\bibinfo{volume}{7902131}},
  \bibinfo{pages}{95} (\bibinfo{year}{1979}).

\bibitem[{\citenamefont{Mohapatra and Senjanovic}(1980)}]{Mohapatra:1979ia}
\bibinfo{author}{\bibfnamefont{R.~N.} \bibnamefont{Mohapatra}}
  \bibnamefont{and}
  \bibinfo{author}{\bibfnamefont{G.}~\bibnamefont{Senjanovic}},
  \bibinfo{journal}{Phys. Rev. Lett.} \textbf{\bibinfo{volume}{44}},
  \bibinfo{pages}{912} (\bibinfo{year}{1980}).

\bibitem[{\citenamefont{Magg and Wetterich}(1980)}]{Magg:1980ut}
\bibinfo{author}{\bibfnamefont{M.}~\bibnamefont{Magg}} \bibnamefont{and}
  \bibinfo{author}{\bibfnamefont{C.}~\bibnamefont{Wetterich}},
  \bibinfo{journal}{Phys. Lett. B} \textbf{\bibinfo{volume}{94}},
  \bibinfo{pages}{61} (\bibinfo{year}{1980}).

\bibitem[{\citenamefont{Schechter and Valle}(1980)}]{Schechter:1980gr}
\bibinfo{author}{\bibfnamefont{J.}~\bibnamefont{Schechter}} \bibnamefont{and}
  \bibinfo{author}{\bibfnamefont{J.}~\bibnamefont{Valle}},
  \bibinfo{journal}{Phys. Rev. D} \textbf{\bibinfo{volume}{22}},
  \bibinfo{pages}{2227} (\bibinfo{year}{1980}).

\bibitem[{\citenamefont{Cheng and Li}(1980)}]{Cheng:1980qt}
\bibinfo{author}{\bibfnamefont{T.}~\bibnamefont{Cheng}} \bibnamefont{and}
  \bibinfo{author}{\bibfnamefont{L.-F.} \bibnamefont{Li}},
  \bibinfo{journal}{Phys. Rev. D} \textbf{\bibinfo{volume}{22}},
  \bibinfo{pages}{2860} (\bibinfo{year}{1980}).

\bibitem[{\citenamefont{Mohapatra and Senjanovic}(1981)}]{Mohapatra:1980yp}
\bibinfo{author}{\bibfnamefont{R.~N.} \bibnamefont{Mohapatra}}
  \bibnamefont{and}
  \bibinfo{author}{\bibfnamefont{G.}~\bibnamefont{Senjanovic}},
  \bibinfo{journal}{Phys. Rev. D} \textbf{\bibinfo{volume}{23}},
  \bibinfo{pages}{165} (\bibinfo{year}{1981}).

\bibitem[{\citenamefont{Foot et~al.}(1989)\citenamefont{Foot, Lew, He, and
  Joshi}}]{Foot:1988aq}
\bibinfo{author}{\bibfnamefont{R.}~\bibnamefont{Foot}},
  \bibinfo{author}{\bibfnamefont{H.}~\bibnamefont{Lew}},
  \bibinfo{author}{\bibfnamefont{X.~G.} \bibnamefont{He}}, \bibnamefont{and}
  \bibinfo{author}{\bibfnamefont{G.~C.} \bibnamefont{Joshi}},
  \bibinfo{journal}{Z. Phys. C} \textbf{\bibinfo{volume}{44}},
  \bibinfo{pages}{441} (\bibinfo{year}{1989}).

\bibitem[{\citenamefont{Ma}(1998)}]{Ma:1998dn}
\bibinfo{author}{\bibfnamefont{E.}~\bibnamefont{Ma}}, \bibinfo{journal}{Phys.
  Rev. Lett.} \textbf{\bibinfo{volume}{81}}, \bibinfo{pages}{1171}
  (\bibinfo{year}{1998}), \eprint{hep-ph/9805219}.

\bibitem[{\citenamefont{Zee}(1980{\natexlab{a}})}]{Zee:1980ai}
\bibinfo{author}{\bibfnamefont{A.}~\bibnamefont{Zee}}, \bibinfo{journal}{Phys.
  Lett. B} \textbf{\bibinfo{volume}{93}}, \bibinfo{pages}{389}
  (\bibinfo{year}{1980}{\natexlab{a}}), \bibinfo{note}{[Erratum: Phys.Lett.B
  95, 461 (1980)]}.

\bibitem[{\citenamefont{Hall and Suzuki}(1984)}]{Hall:1983id}
\bibinfo{author}{\bibfnamefont{L.~J.} \bibnamefont{Hall}} \bibnamefont{and}
  \bibinfo{author}{\bibfnamefont{M.}~\bibnamefont{Suzuki}},
  \bibinfo{journal}{Nucl. Phys. B} \textbf{\bibinfo{volume}{231}},
  \bibinfo{pages}{419} (\bibinfo{year}{1984}).

\bibitem[{\citenamefont{Zee}(1986)}]{Zee:1985id}
\bibinfo{author}{\bibfnamefont{A.}~\bibnamefont{Zee}}, \bibinfo{journal}{Nucl.
  Phys. B} \textbf{\bibinfo{volume}{264}}, \bibinfo{pages}{99}
  (\bibinfo{year}{1986}).

\bibitem[{\citenamefont{Babu}(1988)}]{Babu:1988ki}
\bibinfo{author}{\bibfnamefont{K.~S.} \bibnamefont{Babu}},
  \bibinfo{journal}{Phys. Lett. B} \textbf{\bibinfo{volume}{203}},
  \bibinfo{pages}{132} (\bibinfo{year}{1988}).

\bibitem[{\citenamefont{Krauss et~al.}(2003)\citenamefont{Krauss, Nasri, and
  Trodden}}]{Krauss:2002px}
\bibinfo{author}{\bibfnamefont{L.~M.} \bibnamefont{Krauss}},
  \bibinfo{author}{\bibfnamefont{S.}~\bibnamefont{Nasri}}, \bibnamefont{and}
  \bibinfo{author}{\bibfnamefont{M.}~\bibnamefont{Trodden}},
  \bibinfo{journal}{Phys. Rev. D} \textbf{\bibinfo{volume}{67}},
  \bibinfo{pages}{085002} (\bibinfo{year}{2003}), \eprint{hep-ph/0210389}.

\bibitem[{\citenamefont{Cai et~al.}(2018)\citenamefont{Cai, Han, Li, and
  Ruiz}}]{Cai:2017mow}
\bibinfo{author}{\bibfnamefont{Y.}~\bibnamefont{Cai}},
  \bibinfo{author}{\bibfnamefont{T.}~\bibnamefont{Han}},
  \bibinfo{author}{\bibfnamefont{T.}~\bibnamefont{Li}}, \bibnamefont{and}
  \bibinfo{author}{\bibfnamefont{R.}~\bibnamefont{Ruiz}},
  \bibinfo{journal}{Front. in Phys.} \textbf{\bibinfo{volume}{6}},
  \bibinfo{pages}{40} (\bibinfo{year}{2018}), \eprint{1711.02180}.

\bibitem[{\citenamefont{Babu et~al.}(2020{\natexlab{a}})\citenamefont{Babu,
  Dev, Jana, and Thapa}}]{Babu:2019mfe}
\bibinfo{author}{\bibfnamefont{K.~S.} \bibnamefont{Babu}},
  \bibinfo{author}{\bibfnamefont{P.~S.~B.} \bibnamefont{Dev}},
  \bibinfo{author}{\bibfnamefont{S.}~\bibnamefont{Jana}}, \bibnamefont{and}
  \bibinfo{author}{\bibfnamefont{A.}~\bibnamefont{Thapa}},
  \bibinfo{journal}{JHEP} \textbf{\bibinfo{volume}{03}}, \bibinfo{pages}{006}
  (\bibinfo{year}{2020}{\natexlab{a}}), \eprint{1907.09498}.

\bibitem[{\citenamefont{Fukugita and Yanagida}(1986)}]{Fukugita:1986hr}
\bibinfo{author}{\bibfnamefont{M.}~\bibnamefont{Fukugita}} \bibnamefont{and}
  \bibinfo{author}{\bibfnamefont{T.}~\bibnamefont{Yanagida}},
  \bibinfo{journal}{Phys. Lett. B} \textbf{\bibinfo{volume}{174}},
  \bibinfo{pages}{45} (\bibinfo{year}{1986}).

\bibitem[{\citenamefont{Furry}(1939)}]{PhysRev.56.1184}
\bibinfo{author}{\bibfnamefont{W.~H.} \bibnamefont{Furry}},
  \bibinfo{journal}{Phys. Rev.} \textbf{\bibinfo{volume}{56}},
  \bibinfo{pages}{1184} (\bibinfo{year}{1939}).

\bibitem[{\citenamefont{Rodejohann}(2011)}]{Rodejohann:2011mu}
\bibinfo{author}{\bibfnamefont{W.}~\bibnamefont{Rodejohann}},
  \bibinfo{journal}{Int. J. Mod. Phys. E} \textbf{\bibinfo{volume}{20}},
  \bibinfo{pages}{1833} (\bibinfo{year}{2011}), \eprint{1106.1334}.

\bibitem[{\citenamefont{Keung and Senjanovic}(1983)}]{Keung:1983uu}
\bibinfo{author}{\bibfnamefont{W.-Y.} \bibnamefont{Keung}} \bibnamefont{and}
  \bibinfo{author}{\bibfnamefont{G.}~\bibnamefont{Senjanovic}},
  \bibinfo{journal}{Phys. Rev. Lett.} \textbf{\bibinfo{volume}{50}},
  \bibinfo{pages}{1427} (\bibinfo{year}{1983}).

\bibitem[{\citenamefont{Pati and Salam}(1974)}]{Pati:1974yy}
\bibinfo{author}{\bibfnamefont{J.~C.} \bibnamefont{Pati}} \bibnamefont{and}
  \bibinfo{author}{\bibfnamefont{A.}~\bibnamefont{Salam}},
  \bibinfo{journal}{Phys. Rev. D} \textbf{\bibinfo{volume}{10}},
  \bibinfo{pages}{275} (\bibinfo{year}{1974}), \bibinfo{note}{[Erratum:
  Phys.Rev.D 11, 703--703 (1975)]}.

\bibitem[{\citenamefont{Mohapatra and
  Pati}(1975{\natexlab{a}})}]{Mohapatra:1974hk}
\bibinfo{author}{\bibfnamefont{R.~N.} \bibnamefont{Mohapatra}}
  \bibnamefont{and} \bibinfo{author}{\bibfnamefont{J.~C.} \bibnamefont{Pati}},
  \bibinfo{journal}{Phys. Rev. D} \textbf{\bibinfo{volume}{11}},
  \bibinfo{pages}{566} (\bibinfo{year}{1975}{\natexlab{a}}).

\bibitem[{\citenamefont{Mohapatra and
  Pati}(1975{\natexlab{b}})}]{Mohapatra:1974gc}
\bibinfo{author}{\bibfnamefont{R.~N.} \bibnamefont{Mohapatra}}
  \bibnamefont{and} \bibinfo{author}{\bibfnamefont{J.~C.} \bibnamefont{Pati}},
  \bibinfo{journal}{Phys. Rev. D} \textbf{\bibinfo{volume}{11}},
  \bibinfo{pages}{2558} (\bibinfo{year}{1975}{\natexlab{b}}).

\bibitem[{\citenamefont{Ferrari et~al.}(2000)\citenamefont{Ferrari, Collot,
  Andrieux, Belhorma, de~Saintignon, Hostachy, Martin, and
  Wielers}}]{Ferrari:2000sp}
\bibinfo{author}{\bibfnamefont{A.}~\bibnamefont{Ferrari}},
  \bibinfo{author}{\bibfnamefont{J.}~\bibnamefont{Collot}},
  \bibinfo{author}{\bibfnamefont{M.-L.} \bibnamefont{Andrieux}},
  \bibinfo{author}{\bibfnamefont{B.}~\bibnamefont{Belhorma}},
  \bibinfo{author}{\bibfnamefont{P.}~\bibnamefont{de~Saintignon}},
  \bibinfo{author}{\bibfnamefont{J.-Y.} \bibnamefont{Hostachy}},
  \bibinfo{author}{\bibfnamefont{P.}~\bibnamefont{Martin}}, \bibnamefont{and}
  \bibinfo{author}{\bibfnamefont{M.}~\bibnamefont{Wielers}},
  \bibinfo{journal}{Phys. Rev. D} \textbf{\bibinfo{volume}{62}},
  \bibinfo{pages}{013001} (\bibinfo{year}{2000}).

\bibitem[{\citenamefont{Gninenko et~al.}(2007)\citenamefont{Gninenko, Kirsanov,
  Krasnikov, and Matveev}}]{Gninenko:2006br}
\bibinfo{author}{\bibfnamefont{S.~N.} \bibnamefont{Gninenko}},
  \bibinfo{author}{\bibfnamefont{M.~M.} \bibnamefont{Kirsanov}},
  \bibinfo{author}{\bibfnamefont{N.~V.} \bibnamefont{Krasnikov}},
  \bibnamefont{and} \bibinfo{author}{\bibfnamefont{V.~A.}
  \bibnamefont{Matveev}}, \bibinfo{journal}{Phys. Atom. Nucl.}
  \textbf{\bibinfo{volume}{70}}, \bibinfo{pages}{441} (\bibinfo{year}{2007}).

\bibitem[{\citenamefont{Maiezza et~al.}(2010)\citenamefont{Maiezza, Nemevsek,
  Nesti, and Senjanovic}}]{Maiezza:2010ic}
\bibinfo{author}{\bibfnamefont{A.}~\bibnamefont{Maiezza}},
  \bibinfo{author}{\bibfnamefont{M.}~\bibnamefont{Nemevsek}},
  \bibinfo{author}{\bibfnamefont{F.}~\bibnamefont{Nesti}}, \bibnamefont{and}
  \bibinfo{author}{\bibfnamefont{G.}~\bibnamefont{Senjanovic}},
  \bibinfo{journal}{Phys. Rev. D} \textbf{\bibinfo{volume}{82}},
  \bibinfo{pages}{055022} (\bibinfo{year}{2010}), \eprint{1005.5160}.

\bibitem[{\citenamefont{Nemevsek et~al.}(2011)\citenamefont{Nemevsek, Nesti,
  Senjanovic, and Zhang}}]{Nemevsek:2011hz}
\bibinfo{author}{\bibfnamefont{M.}~\bibnamefont{Nemevsek}},
  \bibinfo{author}{\bibfnamefont{F.}~\bibnamefont{Nesti}},
  \bibinfo{author}{\bibfnamefont{G.}~\bibnamefont{Senjanovic}},
  \bibnamefont{and} \bibinfo{author}{\bibfnamefont{Y.}~\bibnamefont{Zhang}},
  \bibinfo{journal}{Phys. Rev. D} \textbf{\bibinfo{volume}{83}},
  \bibinfo{pages}{115014} (\bibinfo{year}{2011}), \eprint{1103.1627}.

\bibitem[{\citenamefont{Chen and Dev}(2012)}]{Chen:2011hc}
\bibinfo{author}{\bibfnamefont{C.-Y.} \bibnamefont{Chen}} \bibnamefont{and}
  \bibinfo{author}{\bibfnamefont{P.~S.~B.} \bibnamefont{Dev}},
  \bibinfo{journal}{Phys. Rev. D} \textbf{\bibinfo{volume}{85}},
  \bibinfo{pages}{093018} (\bibinfo{year}{2012}), \eprint{1112.6419}.

\bibitem[{\citenamefont{Chakrabortty
  et~al.}(2012{\natexlab{a}})\citenamefont{Chakrabortty, Gluza, Sevillano, and
  Szafron}}]{Chakrabortty:2012pp}
\bibinfo{author}{\bibfnamefont{J.}~\bibnamefont{Chakrabortty}},
  \bibinfo{author}{\bibfnamefont{J.}~\bibnamefont{Gluza}},
  \bibinfo{author}{\bibfnamefont{R.}~\bibnamefont{Sevillano}},
  \bibnamefont{and} \bibinfo{author}{\bibfnamefont{R.}~\bibnamefont{Szafron}},
  \bibinfo{journal}{JHEP} \textbf{\bibinfo{volume}{07}}, \bibinfo{pages}{038}
  (\bibinfo{year}{2012}{\natexlab{a}}), \eprint{1204.0736}.

\bibitem[{\citenamefont{Aguilar-Saavedra and
  Joaquim}(2012)}]{Aguilar-Saavedra:2012grq}
\bibinfo{author}{\bibfnamefont{J.~A.} \bibnamefont{Aguilar-Saavedra}}
  \bibnamefont{and} \bibinfo{author}{\bibfnamefont{F.~R.}
  \bibnamefont{Joaquim}}, \bibinfo{journal}{Phys. Rev. D}
  \textbf{\bibinfo{volume}{86}}, \bibinfo{pages}{073005}
  (\bibinfo{year}{2012}), \eprint{1207.4193}.

\bibitem[{\citenamefont{Han et~al.}(2013)\citenamefont{Han, Lewis, Ruiz, and
  Si}}]{Han:2012vk}
\bibinfo{author}{\bibfnamefont{T.}~\bibnamefont{Han}},
  \bibinfo{author}{\bibfnamefont{I.}~\bibnamefont{Lewis}},
  \bibinfo{author}{\bibfnamefont{R.}~\bibnamefont{Ruiz}}, \bibnamefont{and}
  \bibinfo{author}{\bibfnamefont{Z.-g.} \bibnamefont{Si}},
  \bibinfo{journal}{Phys. Rev. D} \textbf{\bibinfo{volume}{87}},
  \bibinfo{pages}{035011} (\bibinfo{year}{2013}), \bibinfo{note}{[Erratum:
  Phys.Rev.D 87, 039906 (2013)]}, \eprint{1211.6447}.

\bibitem[{\citenamefont{Chen et~al.}(2013)\citenamefont{Chen, Dev, and
  Mohapatra}}]{Chen:2013foz}
\bibinfo{author}{\bibfnamefont{C.-Y.} \bibnamefont{Chen}},
  \bibinfo{author}{\bibfnamefont{P.~S.~B.} \bibnamefont{Dev}},
  \bibnamefont{and} \bibinfo{author}{\bibfnamefont{R.~N.}
  \bibnamefont{Mohapatra}}, \bibinfo{journal}{Phys. Rev. D}
  \textbf{\bibinfo{volume}{88}}, \bibinfo{pages}{033014}
  (\bibinfo{year}{2013}), \eprint{1306.2342}.

\bibitem[{\citenamefont{Dev et~al.}(2014)\citenamefont{Dev, Pilaftsis, and
  Yang}}]{Dev:2013wba}
\bibinfo{author}{\bibfnamefont{P.~S.~B.} \bibnamefont{Dev}},
  \bibinfo{author}{\bibfnamefont{A.}~\bibnamefont{Pilaftsis}},
  \bibnamefont{and} \bibinfo{author}{\bibfnamefont{U.-k.} \bibnamefont{Yang}},
  \bibinfo{journal}{Phys. Rev. Lett.} \textbf{\bibinfo{volume}{112}},
  \bibinfo{pages}{081801} (\bibinfo{year}{2014}), \eprint{1308.2209}.

\bibitem[{\citenamefont{Dutta et~al.}(2014)\citenamefont{Dutta, Eusebi, Gao,
  Ghosh, and Kamon}}]{Dutta:2014dba}
\bibinfo{author}{\bibfnamefont{B.}~\bibnamefont{Dutta}},
  \bibinfo{author}{\bibfnamefont{R.}~\bibnamefont{Eusebi}},
  \bibinfo{author}{\bibfnamefont{Y.}~\bibnamefont{Gao}},
  \bibinfo{author}{\bibfnamefont{T.}~\bibnamefont{Ghosh}}, \bibnamefont{and}
  \bibinfo{author}{\bibfnamefont{T.}~\bibnamefont{Kamon}},
  \bibinfo{journal}{Phys. Rev. D} \textbf{\bibinfo{volume}{90}},
  \bibinfo{pages}{055015} (\bibinfo{year}{2014}), \eprint{1404.0685}.

\bibitem[{\citenamefont{Gluza and Jeli\'nski}(2015)}]{Gluza:2015goa}
\bibinfo{author}{\bibfnamefont{J.}~\bibnamefont{Gluza}} \bibnamefont{and}
  \bibinfo{author}{\bibfnamefont{T.}~\bibnamefont{Jeli\'nski}},
  \bibinfo{journal}{Phys. Lett. B} \textbf{\bibinfo{volume}{748}},
  \bibinfo{pages}{125} (\bibinfo{year}{2015}), \eprint{1504.05568}.

\bibitem[{\citenamefont{Ng et~al.}(2015)\citenamefont{Ng, de~la Puente, and
  Pan}}]{Ng:2015hba}
\bibinfo{author}{\bibfnamefont{J.~N.} \bibnamefont{Ng}},
  \bibinfo{author}{\bibfnamefont{A.}~\bibnamefont{de~la Puente}},
  \bibnamefont{and} \bibinfo{author}{\bibfnamefont{B.~W.-P.}
  \bibnamefont{Pan}}, \bibinfo{journal}{JHEP} \textbf{\bibinfo{volume}{12}},
  \bibinfo{pages}{172} (\bibinfo{year}{2015}), \eprint{1505.01934}.

\bibitem[{\citenamefont{Maiezza et~al.}(2015)\citenamefont{Maiezza,
  Nemev\v{s}ek, and Nesti}}]{Maiezza:2015lza}
\bibinfo{author}{\bibfnamefont{A.}~\bibnamefont{Maiezza}},
  \bibinfo{author}{\bibfnamefont{M.}~\bibnamefont{Nemev\v{s}ek}},
  \bibnamefont{and} \bibinfo{author}{\bibfnamefont{F.}~\bibnamefont{Nesti}},
  \bibinfo{journal}{Phys. Rev. Lett.} \textbf{\bibinfo{volume}{115}},
  \bibinfo{pages}{081802} (\bibinfo{year}{2015}), \eprint{1503.06834}.

\bibitem[{\citenamefont{Deppisch et~al.}(2015)\citenamefont{Deppisch,
  Bhupal~Dev, and Pilaftsis}}]{Deppisch:2015qwa}
\bibinfo{author}{\bibfnamefont{F.~F.} \bibnamefont{Deppisch}},
  \bibinfo{author}{\bibfnamefont{P.~S.} \bibnamefont{Bhupal~Dev}},
  \bibnamefont{and}
  \bibinfo{author}{\bibfnamefont{A.}~\bibnamefont{Pilaftsis}},
  \bibinfo{journal}{New J. Phys.} \textbf{\bibinfo{volume}{17}},
  \bibinfo{pages}{075019} (\bibinfo{year}{2015}), \eprint{1502.06541}.

\bibitem[{\citenamefont{Degrande et~al.}(2016)\citenamefont{Degrande,
  Mattelaer, Ruiz, and Turner}}]{Degrande:2016aje}
\bibinfo{author}{\bibfnamefont{C.}~\bibnamefont{Degrande}},
  \bibinfo{author}{\bibfnamefont{O.}~\bibnamefont{Mattelaer}},
  \bibinfo{author}{\bibfnamefont{R.}~\bibnamefont{Ruiz}}, \bibnamefont{and}
  \bibinfo{author}{\bibfnamefont{J.}~\bibnamefont{Turner}},
  \bibinfo{journal}{Phys. Rev. D} \textbf{\bibinfo{volume}{94}},
  \bibinfo{pages}{053002} (\bibinfo{year}{2016}), \eprint{1602.06957}.

\bibitem[{\citenamefont{Dev et~al.}(2016)\citenamefont{Dev, Mohapatra, and
  Zhang}}]{Dev:2016dja}
\bibinfo{author}{\bibfnamefont{P.~S.~B.} \bibnamefont{Dev}},
  \bibinfo{author}{\bibfnamefont{R.~N.} \bibnamefont{Mohapatra}},
  \bibnamefont{and} \bibinfo{author}{\bibfnamefont{Y.}~\bibnamefont{Zhang}},
  \bibinfo{journal}{JHEP} \textbf{\bibinfo{volume}{05}}, \bibinfo{pages}{174}
  (\bibinfo{year}{2016}), \eprint{1602.05947}.

\bibitem[{\citenamefont{Roitgrund}(2017)}]{Roitgrund:2017byx}
\bibinfo{author}{\bibfnamefont{A.}~\bibnamefont{Roitgrund}}
  (\bibinfo{year}{2017}), \eprint{1704.07772}.

\bibitem[{\citenamefont{Nemev\v{s}ek et~al.}(2018)\citenamefont{Nemev\v{s}ek,
  Nesti, and Popara}}]{Nemevsek:2018bbt}
\bibinfo{author}{\bibfnamefont{M.}~\bibnamefont{Nemev\v{s}ek}},
  \bibinfo{author}{\bibfnamefont{F.}~\bibnamefont{Nesti}}, \bibnamefont{and}
  \bibinfo{author}{\bibfnamefont{G.}~\bibnamefont{Popara}},
  \bibinfo{journal}{Phys. Rev. D} \textbf{\bibinfo{volume}{97}},
  \bibinfo{pages}{115018} (\bibinfo{year}{2018}), \eprint{1801.05813}.

\bibitem[{\citenamefont{Dicus et~al.}(1991)\citenamefont{Dicus, Karatas, and
  Roy}}]{Dicus:1991fk}
\bibinfo{author}{\bibfnamefont{D.~A.} \bibnamefont{Dicus}},
  \bibinfo{author}{\bibfnamefont{D.~D.} \bibnamefont{Karatas}},
  \bibnamefont{and} \bibinfo{author}{\bibfnamefont{P.}~\bibnamefont{Roy}},
  \bibinfo{journal}{Phys. Rev. D} \textbf{\bibinfo{volume}{44}},
  \bibinfo{pages}{2033} (\bibinfo{year}{1991}).

\bibitem[{\citenamefont{Datta et~al.}(1994)\citenamefont{Datta, Guchait, and
  Pilaftsis}}]{Datta:1993nm}
\bibinfo{author}{\bibfnamefont{A.}~\bibnamefont{Datta}},
  \bibinfo{author}{\bibfnamefont{M.}~\bibnamefont{Guchait}}, \bibnamefont{and}
  \bibinfo{author}{\bibfnamefont{A.}~\bibnamefont{Pilaftsis}},
  \bibinfo{journal}{Phys. Rev. D} \textbf{\bibinfo{volume}{50}},
  \bibinfo{pages}{3195} (\bibinfo{year}{1994}), \eprint{hep-ph/9311257}.

\bibitem[{\citenamefont{Ali et~al.}(2001)\citenamefont{Ali, Borisov, and
  Zamorin}}]{Ali:2001gsa}
\bibinfo{author}{\bibfnamefont{A.}~\bibnamefont{Ali}},
  \bibinfo{author}{\bibfnamefont{A.~V.} \bibnamefont{Borisov}},
  \bibnamefont{and} \bibinfo{author}{\bibfnamefont{N.~B.}
  \bibnamefont{Zamorin}}, \bibinfo{journal}{Eur. Phys. J. C}
  \textbf{\bibinfo{volume}{21}}, \bibinfo{pages}{123} (\bibinfo{year}{2001}),
  \eprint{hep-ph/0104123}.

\bibitem[{\citenamefont{Han and Zhang}(2006)}]{Han:2006ip}
\bibinfo{author}{\bibfnamefont{T.}~\bibnamefont{Han}} \bibnamefont{and}
  \bibinfo{author}{\bibfnamefont{B.}~\bibnamefont{Zhang}},
  \bibinfo{journal}{Phys. Rev. Lett.} \textbf{\bibinfo{volume}{97}},
  \bibinfo{pages}{171804} (\bibinfo{year}{2006}), \eprint{hep-ph/0604064}.

\bibitem[{\citenamefont{Kersten and Smirnov}(2007)}]{Kersten:2007vk}
\bibinfo{author}{\bibfnamefont{J.}~\bibnamefont{Kersten}} \bibnamefont{and}
  \bibinfo{author}{\bibfnamefont{A.~Y.} \bibnamefont{Smirnov}},
  \bibinfo{journal}{Phys. Rev. D} \textbf{\bibinfo{volume}{76}},
  \bibinfo{pages}{073005} (\bibinfo{year}{2007}), \eprint{0705.3221}.

\bibitem[{\citenamefont{del Aguila et~al.}(2007)\citenamefont{del Aguila,
  Aguilar-Saavedra, and Pittau}}]{delAguila:2007qnc}
\bibinfo{author}{\bibfnamefont{F.}~\bibnamefont{del Aguila}},
  \bibinfo{author}{\bibfnamefont{J.~A.} \bibnamefont{Aguilar-Saavedra}},
  \bibnamefont{and} \bibinfo{author}{\bibfnamefont{R.}~\bibnamefont{Pittau}},
  \bibinfo{journal}{JHEP} \textbf{\bibinfo{volume}{10}}, \bibinfo{pages}{047}
  (\bibinfo{year}{2007}), \eprint{hep-ph/0703261}.

\bibitem[{\citenamefont{Atre et~al.}(2009)\citenamefont{Atre, Han, Pascoli, and
  Zhang}}]{Atre:2009rg}
\bibinfo{author}{\bibfnamefont{A.}~\bibnamefont{Atre}},
  \bibinfo{author}{\bibfnamefont{T.}~\bibnamefont{Han}},
  \bibinfo{author}{\bibfnamefont{S.}~\bibnamefont{Pascoli}}, \bibnamefont{and}
  \bibinfo{author}{\bibfnamefont{B.}~\bibnamefont{Zhang}},
  \bibinfo{journal}{JHEP} \textbf{\bibinfo{volume}{05}}, \bibinfo{pages}{030}
  (\bibinfo{year}{2009}), \eprint{0901.3589}.

\bibitem[{\citenamefont{Alva et~al.}(2015)\citenamefont{Alva, Han, and
  Ruiz}}]{Alva:2014gxa}
\bibinfo{author}{\bibfnamefont{D.}~\bibnamefont{Alva}},
  \bibinfo{author}{\bibfnamefont{T.}~\bibnamefont{Han}}, \bibnamefont{and}
  \bibinfo{author}{\bibfnamefont{R.}~\bibnamefont{Ruiz}},
  \bibinfo{journal}{JHEP} \textbf{\bibinfo{volume}{02}}, \bibinfo{pages}{072}
  (\bibinfo{year}{2015}), \eprint{1411.7305}.

\bibitem[{\citenamefont{Das and Okada}(2016)}]{Das:2015toa}
\bibinfo{author}{\bibfnamefont{A.}~\bibnamefont{Das}} \bibnamefont{and}
  \bibinfo{author}{\bibfnamefont{N.}~\bibnamefont{Okada}},
  \bibinfo{journal}{Phys. Rev. D} \textbf{\bibinfo{volume}{93}},
  \bibinfo{pages}{033003} (\bibinfo{year}{2016}), \eprint{1510.04790}.

\bibitem[{\citenamefont{Drewes et~al.}(2019)\citenamefont{Drewes, Klari\'c, and
  Klose}}]{Drewes:2019byd}
\bibinfo{author}{\bibfnamefont{M.}~\bibnamefont{Drewes}},
  \bibinfo{author}{\bibfnamefont{J.}~\bibnamefont{Klari\'c}}, \bibnamefont{and}
  \bibinfo{author}{\bibfnamefont{P.}~\bibnamefont{Klose}},
  \bibinfo{journal}{JHEP} \textbf{\bibinfo{volume}{11}}, \bibinfo{pages}{032}
  (\bibinfo{year}{2019}), \eprint{1907.13034}.

\bibitem[{\citenamefont{Fuks et~al.}(2021{\natexlab{a}})\citenamefont{Fuks,
  Neundorf, Peters, Ruiz, and Saimpert}}]{Fuks:2020att}
\bibinfo{author}{\bibfnamefont{B.}~\bibnamefont{Fuks}},
  \bibinfo{author}{\bibfnamefont{J.}~\bibnamefont{Neundorf}},
  \bibinfo{author}{\bibfnamefont{K.}~\bibnamefont{Peters}},
  \bibinfo{author}{\bibfnamefont{R.}~\bibnamefont{Ruiz}}, \bibnamefont{and}
  \bibinfo{author}{\bibfnamefont{M.}~\bibnamefont{Saimpert}},
  \bibinfo{journal}{Phys. Rev. D} \textbf{\bibinfo{volume}{103}},
  \bibinfo{pages}{055005} (\bibinfo{year}{2021}{\natexlab{a}}),
  \eprint{2011.02547}.

\bibitem[{\citenamefont{Akeroyd et~al.}(2008)\citenamefont{Akeroyd, Aoki, and
  Sugiyama}}]{Akeroyd:2007zv}
\bibinfo{author}{\bibfnamefont{A.~G.} \bibnamefont{Akeroyd}},
  \bibinfo{author}{\bibfnamefont{M.}~\bibnamefont{Aoki}}, \bibnamefont{and}
  \bibinfo{author}{\bibfnamefont{H.}~\bibnamefont{Sugiyama}},
  \bibinfo{journal}{Phys. Rev. D} \textbf{\bibinfo{volume}{77}},
  \bibinfo{pages}{075010} (\bibinfo{year}{2008}), \eprint{0712.4019}.

\bibitem[{\citenamefont{Fileviez~Perez
  et~al.}(2008{\natexlab{a}})\citenamefont{Fileviez~Perez, Han, Huang, Li, and
  Wang}}]{Perez:2008zc}
\bibinfo{author}{\bibfnamefont{P.}~\bibnamefont{Fileviez~Perez}},
  \bibinfo{author}{\bibfnamefont{T.}~\bibnamefont{Han}},
  \bibinfo{author}{\bibfnamefont{G.-Y.} \bibnamefont{Huang}},
  \bibinfo{author}{\bibfnamefont{T.}~\bibnamefont{Li}}, \bibnamefont{and}
  \bibinfo{author}{\bibfnamefont{K.}~\bibnamefont{Wang}},
  \bibinfo{journal}{Phys. Rev. D} \textbf{\bibinfo{volume}{78}},
  \bibinfo{pages}{071301} (\bibinfo{year}{2008}{\natexlab{a}}),
  \eprint{0803.3450}.

\bibitem[{\citenamefont{Fileviez~Perez
  et~al.}(2008{\natexlab{b}})\citenamefont{Fileviez~Perez, Han, Huang, Li, and
  Wang}}]{FileviezPerez:2008jbu}
\bibinfo{author}{\bibfnamefont{P.}~\bibnamefont{Fileviez~Perez}},
  \bibinfo{author}{\bibfnamefont{T.}~\bibnamefont{Han}},
  \bibinfo{author}{\bibfnamefont{G.-y.} \bibnamefont{Huang}},
  \bibinfo{author}{\bibfnamefont{T.}~\bibnamefont{Li}}, \bibnamefont{and}
  \bibinfo{author}{\bibfnamefont{K.}~\bibnamefont{Wang}},
  \bibinfo{journal}{Phys. Rev. D} \textbf{\bibinfo{volume}{78}},
  \bibinfo{pages}{015018} (\bibinfo{year}{2008}{\natexlab{b}}),
  \eprint{0805.3536}.

\bibitem[{\citenamefont{Melfo et~al.}(2012)\citenamefont{Melfo, Nemevsek,
  Nesti, Senjanovic, and Zhang}}]{Melfo:2011nx}
\bibinfo{author}{\bibfnamefont{A.}~\bibnamefont{Melfo}},
  \bibinfo{author}{\bibfnamefont{M.}~\bibnamefont{Nemevsek}},
  \bibinfo{author}{\bibfnamefont{F.}~\bibnamefont{Nesti}},
  \bibinfo{author}{\bibfnamefont{G.}~\bibnamefont{Senjanovic}},
  \bibnamefont{and} \bibinfo{author}{\bibfnamefont{Y.}~\bibnamefont{Zhang}},
  \bibinfo{journal}{Phys. Rev. D} \textbf{\bibinfo{volume}{85}},
  \bibinfo{pages}{055018} (\bibinfo{year}{2012}), \eprint{1108.4416}.

\bibitem[{\citenamefont{Aaboud et~al.}(2018{\natexlab{a}})}]{Aaboud:2017qph}
\bibinfo{author}{\bibfnamefont{M.}~\bibnamefont{Aaboud}} \bibnamefont{et~al.}
  (\bibinfo{collaboration}{ATLAS}), \bibinfo{journal}{Eur. Phys. J. C}
  \textbf{\bibinfo{volume}{78}}, \bibinfo{pages}{199}
  (\bibinfo{year}{2018}{\natexlab{a}}), \eprint{1710.09748}.

\bibitem[{\citenamefont{{CMS Collaboration}}(2017)}]{CMS:2017pet}
\bibinfo{author}{\bibnamefont{{CMS Collaboration}}} (\bibinfo{year}{2017}),
  \eprint{CMS-PAS-HIG-16-036}.

\bibitem[{\citenamefont{{ATLAS Collaboration}}(2022)}]{ATLAS:2022pbd}
\bibinfo{author}{\bibnamefont{{ATLAS Collaboration}}} (\bibinfo{year}{2022}),
  \eprint{2211.07505}.

\bibitem[{\citenamefont{Quintero}(2013)}]{Quintero:2012jy}
\bibinfo{author}{\bibfnamefont{N.}~\bibnamefont{Quintero}},
  \bibinfo{journal}{Phys. Rev. D} \textbf{\bibinfo{volume}{87}},
  \bibinfo{pages}{056005} (\bibinfo{year}{2013}), \eprint{1212.3016}.

\bibitem[{\citenamefont{Fuks et~al.}(2021{\natexlab{b}})\citenamefont{Fuks,
  Neundorf, Peters, Ruiz, and Saimpert}}]{Fuks:2020zbm}
\bibinfo{author}{\bibfnamefont{B.}~\bibnamefont{Fuks}},
  \bibinfo{author}{\bibfnamefont{J.}~\bibnamefont{Neundorf}},
  \bibinfo{author}{\bibfnamefont{K.}~\bibnamefont{Peters}},
  \bibinfo{author}{\bibfnamefont{R.}~\bibnamefont{Ruiz}}, \bibnamefont{and}
  \bibinfo{author}{\bibfnamefont{M.}~\bibnamefont{Saimpert}},
  \bibinfo{journal}{Phys. Rev. D} \textbf{\bibinfo{volume}{103}},
  \bibinfo{pages}{115014} (\bibinfo{year}{2021}{\natexlab{b}}),
  \eprint{2012.09882}.

\bibitem[{\citenamefont{Aoki et~al.}(2020)\citenamefont{Aoki, Enomoto, and
  Kanemura}}]{Aoki:2020til}
\bibinfo{author}{\bibfnamefont{M.}~\bibnamefont{Aoki}},
  \bibinfo{author}{\bibfnamefont{K.}~\bibnamefont{Enomoto}}, \bibnamefont{and}
  \bibinfo{author}{\bibfnamefont{S.}~\bibnamefont{Kanemura}},
  \bibinfo{journal}{Phys. Rev. D} \textbf{\bibinfo{volume}{101}},
  \bibinfo{pages}{115019} (\bibinfo{year}{2020}), \eprint{2002.12265}.

\bibitem[{\citenamefont{Harz et~al.}(2021)\citenamefont{Harz, Ramsey-Musolf,
  Shen, and Urrutia-Quiroga}}]{Harz:2021psp}
\bibinfo{author}{\bibfnamefont{J.}~\bibnamefont{Harz}},
  \bibinfo{author}{\bibfnamefont{M.~J.} \bibnamefont{Ramsey-Musolf}},
  \bibinfo{author}{\bibfnamefont{T.}~\bibnamefont{Shen}}, \bibnamefont{and}
  \bibinfo{author}{\bibfnamefont{S.}~\bibnamefont{Urrutia-Quiroga}}
  (\bibinfo{year}{2021}), \eprint{2106.10838}.

\bibitem[{\citenamefont{Graesser et~al.}(2022)\citenamefont{Graesser, Li,
  Ramsey-Musolf, Shen, and Urrutia-Quiroga}}]{Graesser:2022nkv}
\bibinfo{author}{\bibfnamefont{M.~L.} \bibnamefont{Graesser}},
  \bibinfo{author}{\bibfnamefont{G.}~\bibnamefont{Li}},
  \bibinfo{author}{\bibfnamefont{M.~J.} \bibnamefont{Ramsey-Musolf}},
  \bibinfo{author}{\bibfnamefont{T.}~\bibnamefont{Shen}}, \bibnamefont{and}
  \bibinfo{author}{\bibfnamefont{S.}~\bibnamefont{Urrutia-Quiroga}},
  \bibinfo{journal}{JHEP} \textbf{\bibinfo{volume}{10}}, \bibinfo{pages}{034}
  (\bibinfo{year}{2022}), \eprint{2202.01237}.

\bibitem[{\citenamefont{Helo et~al.}(2013{\natexlab{a}})\citenamefont{Helo,
  Hirsch, Kovalenko, and Pas}}]{Helo:2013dla}
\bibinfo{author}{\bibfnamefont{J.~C.} \bibnamefont{Helo}},
  \bibinfo{author}{\bibfnamefont{M.}~\bibnamefont{Hirsch}},
  \bibinfo{author}{\bibfnamefont{S.~G.} \bibnamefont{Kovalenko}},
  \bibnamefont{and} \bibinfo{author}{\bibfnamefont{H.}~\bibnamefont{Pas}},
  \bibinfo{journal}{Phys. Rev. D} \textbf{\bibinfo{volume}{88}},
  \bibinfo{pages}{011901} (\bibinfo{year}{2013}{\natexlab{a}}),
  \eprint{1303.0899}.

\bibitem[{\citenamefont{Helo et~al.}(2013{\natexlab{b}})\citenamefont{Helo,
  Hirsch, P\"as, and Kovalenko}}]{Helo:2013ika}
\bibinfo{author}{\bibfnamefont{J.~C.} \bibnamefont{Helo}},
  \bibinfo{author}{\bibfnamefont{M.}~\bibnamefont{Hirsch}},
  \bibinfo{author}{\bibfnamefont{H.}~\bibnamefont{P\"as}}, \bibnamefont{and}
  \bibinfo{author}{\bibfnamefont{S.~G.} \bibnamefont{Kovalenko}},
  \bibinfo{journal}{Phys. Rev. D} \textbf{\bibinfo{volume}{88}},
  \bibinfo{pages}{073011} (\bibinfo{year}{2013}{\natexlab{b}}),
  \eprint{1307.4849}.

\bibitem[{\citenamefont{del Aguila et~al.}(2013)\citenamefont{del Aguila,
  Chala, Santamaria, and Wudka}}]{delAguila:2013yaa}
\bibinfo{author}{\bibfnamefont{F.}~\bibnamefont{del Aguila}},
  \bibinfo{author}{\bibfnamefont{M.}~\bibnamefont{Chala}},
  \bibinfo{author}{\bibfnamefont{A.}~\bibnamefont{Santamaria}},
  \bibnamefont{and} \bibinfo{author}{\bibfnamefont{J.}~\bibnamefont{Wudka}},
  \bibinfo{journal}{Phys. Lett. B} \textbf{\bibinfo{volume}{725}},
  \bibinfo{pages}{310} (\bibinfo{year}{2013}), \eprint{1305.3904}.

\bibitem[{\citenamefont{del \'Aguila and Chala}(2014)}]{delAguila:2013mia}
\bibinfo{author}{\bibfnamefont{F.}~\bibnamefont{del \'Aguila}}
  \bibnamefont{and} \bibinfo{author}{\bibfnamefont{M.}~\bibnamefont{Chala}},
  \bibinfo{journal}{JHEP} \textbf{\bibinfo{volume}{03}}, \bibinfo{pages}{027}
  (\bibinfo{year}{2014}), \eprint{1311.1510}.

\bibitem[{\citenamefont{Peng et~al.}(2016)\citenamefont{Peng, Ramsey-Musolf,
  and Winslow}}]{Peng:2015haa}
\bibinfo{author}{\bibfnamefont{T.}~\bibnamefont{Peng}},
  \bibinfo{author}{\bibfnamefont{M.~J.} \bibnamefont{Ramsey-Musolf}},
  \bibnamefont{and} \bibinfo{author}{\bibfnamefont{P.}~\bibnamefont{Winslow}},
  \bibinfo{journal}{Phys. Rev. D} \textbf{\bibinfo{volume}{93}},
  \bibinfo{pages}{093002} (\bibinfo{year}{2016}), \eprint{1508.04444}.

\bibitem[{\citenamefont{Cepedello et~al.}(2018)\citenamefont{Cepedello, Hirsch,
  and Helo}}]{Cepedello:2017lyo}
\bibinfo{author}{\bibfnamefont{R.}~\bibnamefont{Cepedello}},
  \bibinfo{author}{\bibfnamefont{M.}~\bibnamefont{Hirsch}}, \bibnamefont{and}
  \bibinfo{author}{\bibfnamefont{J.~C.} \bibnamefont{Helo}},
  \bibinfo{journal}{JHEP} \textbf{\bibinfo{volume}{01}}, \bibinfo{pages}{009}
  (\bibinfo{year}{2018}), \eprint{1709.03397}.

\bibitem[{\citenamefont{Carquin et~al.}(2019)\citenamefont{Carquin, Neill,
  Helo, and Hirsch}}]{Carquin:2019xiz}
\bibinfo{author}{\bibfnamefont{E.}~\bibnamefont{Carquin}},
  \bibinfo{author}{\bibfnamefont{N.~A.} \bibnamefont{Neill}},
  \bibinfo{author}{\bibfnamefont{J.~C.} \bibnamefont{Helo}}, \bibnamefont{and}
  \bibinfo{author}{\bibfnamefont{M.}~\bibnamefont{Hirsch}},
  \bibinfo{journal}{Phys. Rev. D} \textbf{\bibinfo{volume}{99}},
  \bibinfo{pages}{115028} (\bibinfo{year}{2019}), \eprint{1904.07257}.

\bibitem[{\citenamefont{Aristizabal~Sierra and
  Restrepo}(2006)}]{AristizabalSierra:2006ri}
\bibinfo{author}{\bibfnamefont{D.}~\bibnamefont{Aristizabal~Sierra}}
  \bibnamefont{and} \bibinfo{author}{\bibfnamefont{D.}~\bibnamefont{Restrepo}},
  \bibinfo{journal}{JHEP} \textbf{\bibinfo{volume}{08}}, \bibinfo{pages}{036}
  (\bibinfo{year}{2006}), \eprint{hep-ph/0604012}.

\bibitem[{\citenamefont{Herrero-Garc\'\i{}a
  et~al.}(2017)\citenamefont{Herrero-Garc\'\i{}a, Ohlsson, Riad, and
  Wir\'en}}]{Herrero-Garcia:2017xdu}
\bibinfo{author}{\bibfnamefont{J.}~\bibnamefont{Herrero-Garc\'\i{}a}},
  \bibinfo{author}{\bibfnamefont{T.}~\bibnamefont{Ohlsson}},
  \bibinfo{author}{\bibfnamefont{S.}~\bibnamefont{Riad}}, \bibnamefont{and}
  \bibinfo{author}{\bibfnamefont{J.}~\bibnamefont{Wir\'en}},
  \bibinfo{journal}{JHEP} \textbf{\bibinfo{volume}{04}}, \bibinfo{pages}{130}
  (\bibinfo{year}{2017}), \eprint{1701.05345}.

\bibitem[{\citenamefont{Babu et~al.}(2020{\natexlab{b}})\citenamefont{Babu,
  Jana, and Lindner}}]{Babu:2020ivd}
\bibinfo{author}{\bibfnamefont{K.~S.} \bibnamefont{Babu}},
  \bibinfo{author}{\bibfnamefont{S.}~\bibnamefont{Jana}}, \bibnamefont{and}
  \bibinfo{author}{\bibfnamefont{M.}~\bibnamefont{Lindner}},
  \bibinfo{journal}{JHEP} \textbf{\bibinfo{volume}{10}}, \bibinfo{pages}{040}
  (\bibinfo{year}{2020}{\natexlab{b}}), \eprint{2007.04291}.

\bibitem[{\citenamefont{Barman et~al.}(2022)\citenamefont{Barman, Dcruz, and
  Thapa}}]{Barman:2021xeq}
\bibinfo{author}{\bibfnamefont{R.~K.} \bibnamefont{Barman}},
  \bibinfo{author}{\bibfnamefont{R.}~\bibnamefont{Dcruz}}, \bibnamefont{and}
  \bibinfo{author}{\bibfnamefont{A.}~\bibnamefont{Thapa}},
  \bibinfo{journal}{JHEP} \textbf{\bibinfo{volume}{03}}, \bibinfo{pages}{183}
  (\bibinfo{year}{2022}), \eprint{2112.04523}.

\bibitem[{\citenamefont{Schechter and Valle}(1982)}]{Schechter:1981bd}
\bibinfo{author}{\bibfnamefont{J.}~\bibnamefont{Schechter}} \bibnamefont{and}
  \bibinfo{author}{\bibfnamefont{J.}~\bibnamefont{Valle}},
  \bibinfo{journal}{Phys. Rev. D} \textbf{\bibinfo{volume}{25}},
  \bibinfo{pages}{2951} (\bibinfo{year}{1982}).

\bibitem[{\citenamefont{Duerr et~al.}(2011)\citenamefont{Duerr, Lindner, and
  Merle}}]{Duerr:2011zd}
\bibinfo{author}{\bibfnamefont{M.}~\bibnamefont{Duerr}},
  \bibinfo{author}{\bibfnamefont{M.}~\bibnamefont{Lindner}}, \bibnamefont{and}
  \bibinfo{author}{\bibfnamefont{A.}~\bibnamefont{Merle}},
  \bibinfo{journal}{JHEP} \textbf{\bibinfo{volume}{06}}, \bibinfo{pages}{091}
  (\bibinfo{year}{2011}), \eprint{1105.0901}.

\bibitem[{\citenamefont{Agostini et~al.}(2020)}]{GERDA:2020xhi}
\bibinfo{author}{\bibfnamefont{M.}~\bibnamefont{Agostini}} \bibnamefont{et~al.}
  (\bibinfo{collaboration}{GERDA}), \bibinfo{journal}{Phys. Rev. Lett.}
  \textbf{\bibinfo{volume}{125}}, \bibinfo{pages}{252502}
  (\bibinfo{year}{2020}), \eprint{2009.06079}.

\bibitem[{\citenamefont{Klapdor-Kleingrothaus
  et~al.}(2001)}]{Klapdor-Kleingrothaus:2000eir}
\bibinfo{author}{\bibfnamefont{H.~V.} \bibnamefont{Klapdor-Kleingrothaus}}
  \bibnamefont{et~al.}, \bibinfo{journal}{Eur. Phys. J. A}
  \textbf{\bibinfo{volume}{12}}, \bibinfo{pages}{147} (\bibinfo{year}{2001}),
  \eprint{hep-ph/0103062}.

\bibitem[{\citenamefont{Auger et~al.}(2012)}]{EXO-200:2012pdt}
\bibinfo{author}{\bibfnamefont{M.}~\bibnamefont{Auger}} \bibnamefont{et~al.}
  (\bibinfo{collaboration}{EXO-200}), \bibinfo{journal}{Phys. Rev. Lett.}
  \textbf{\bibinfo{volume}{109}}, \bibinfo{pages}{032505}
  (\bibinfo{year}{2012}), \eprint{1205.5608}.

\bibitem[{\citenamefont{Alenkov et~al.}(2019)}]{Alenkov:2019jis}
\bibinfo{author}{\bibfnamefont{V.}~\bibnamefont{Alenkov}} \bibnamefont{et~al.},
  \bibinfo{journal}{Eur. Phys. J. C} \textbf{\bibinfo{volume}{79}},
  \bibinfo{pages}{791} (\bibinfo{year}{2019}), \eprint{1903.09483}.

\bibitem[{\citenamefont{Abgrall et~al.}(2021)}]{LEGEND:2021bnm}
\bibinfo{author}{\bibfnamefont{N.}~\bibnamefont{Abgrall}} \bibnamefont{et~al.}
  (\bibinfo{collaboration}{LEGEND}) (\bibinfo{year}{2021}),
  \eprint{2107.11462}.

\bibitem[{\citenamefont{Chakrabortty
  et~al.}(2012{\natexlab{b}})\citenamefont{Chakrabortty, Devi, Goswami, and
  Patra}}]{Chakrabortty:2012mh}
\bibinfo{author}{\bibfnamefont{J.}~\bibnamefont{Chakrabortty}},
  \bibinfo{author}{\bibfnamefont{H.~Z.} \bibnamefont{Devi}},
  \bibinfo{author}{\bibfnamefont{S.}~\bibnamefont{Goswami}}, \bibnamefont{and}
  \bibinfo{author}{\bibfnamefont{S.}~\bibnamefont{Patra}},
  \bibinfo{journal}{JHEP} \textbf{\bibinfo{volume}{08}}, \bibinfo{pages}{008}
  (\bibinfo{year}{2012}{\natexlab{b}}), \eprint{1204.2527}.

\bibitem[{\citenamefont{Dev et~al.}(2018)\citenamefont{Dev, Ramsey-Musolf, and
  Zhang}}]{Dev:2018sel}
\bibinfo{author}{\bibfnamefont{P.~S.~B.} \bibnamefont{Dev}},
  \bibinfo{author}{\bibfnamefont{M.~J.} \bibnamefont{Ramsey-Musolf}},
  \bibnamefont{and} \bibinfo{author}{\bibfnamefont{Y.}~\bibnamefont{Zhang}},
  \bibinfo{journal}{Phys. Rev. D} \textbf{\bibinfo{volume}{98}},
  \bibinfo{pages}{055013} (\bibinfo{year}{2018}), \eprint{1806.08499}.

\bibitem[{\citenamefont{Kanemura and Yagyu}(2012)}]{Kanemura:2012rs}
\bibinfo{author}{\bibfnamefont{S.}~\bibnamefont{Kanemura}} \bibnamefont{and}
  \bibinfo{author}{\bibfnamefont{K.}~\bibnamefont{Yagyu}},
  \bibinfo{journal}{Phys. Rev. D} \textbf{\bibinfo{volume}{85}},
  \bibinfo{pages}{115009} (\bibinfo{year}{2012}), \eprint{1201.6287}.

\bibitem[{\citenamefont{Alwall et~al.}(2014)\citenamefont{Alwall, Frederix,
  Frixione, Hirschi, Maltoni, Mattelaer, Shao, Stelzer, Torrielli, and
  Zaro}}]{Alwall:2014hca}
\bibinfo{author}{\bibfnamefont{J.}~\bibnamefont{Alwall}},
  \bibinfo{author}{\bibfnamefont{R.}~\bibnamefont{Frederix}},
  \bibinfo{author}{\bibfnamefont{S.}~\bibnamefont{Frixione}},
  \bibinfo{author}{\bibfnamefont{V.}~\bibnamefont{Hirschi}},
  \bibinfo{author}{\bibfnamefont{F.}~\bibnamefont{Maltoni}},
  \bibinfo{author}{\bibfnamefont{O.}~\bibnamefont{Mattelaer}},
  \bibinfo{author}{\bibfnamefont{H.~S.} \bibnamefont{Shao}},
  \bibinfo{author}{\bibfnamefont{T.}~\bibnamefont{Stelzer}},
  \bibinfo{author}{\bibfnamefont{P.}~\bibnamefont{Torrielli}},
  \bibnamefont{and} \bibinfo{author}{\bibfnamefont{M.}~\bibnamefont{Zaro}},
  \bibinfo{journal}{JHEP} \textbf{\bibinfo{volume}{07}}, \bibinfo{pages}{079}
  (\bibinfo{year}{2014}), \eprint{1405.0301}.

\bibitem[{\citenamefont{Fuks et~al.}(2020)\citenamefont{Fuks, Nemev\v~sek, and
  Ruiz}}]{Fuks:2019clu}
\bibinfo{author}{\bibfnamefont{B.}~\bibnamefont{Fuks}},
  \bibinfo{author}{\bibfnamefont{M.}~\bibnamefont{Nemev\v~sek}},
  \bibnamefont{and} \bibinfo{author}{\bibfnamefont{R.}~\bibnamefont{Ruiz}},
  \bibinfo{journal}{Phys. Rev. D} \textbf{\bibinfo{volume}{101}},
  \bibinfo{pages}{075022} (\bibinfo{year}{2020}), \eprint{1912.08975}.

\bibitem[{\citenamefont{Sjostrand et~al.}(2008)\citenamefont{Sjostrand, Mrenna,
  and Skands}}]{Sjostrand:2007gs}
\bibinfo{author}{\bibfnamefont{T.}~\bibnamefont{Sjostrand}},
  \bibinfo{author}{\bibfnamefont{S.}~\bibnamefont{Mrenna}}, \bibnamefont{and}
  \bibinfo{author}{\bibfnamefont{P.~Z.} \bibnamefont{Skands}},
  \bibinfo{journal}{Comput. Phys. Commun.} \textbf{\bibinfo{volume}{178}},
  \bibinfo{pages}{852} (\bibinfo{year}{2008}), \eprint{0710.3820}.

\bibitem[{\citenamefont{de~Favereau et~al.}(2014)\citenamefont{de~Favereau,
  Delaere, Demin, Giammanco, Lemaître, Mertens, and
  Selvaggi}}]{deFavereau:2013fsa}
\bibinfo{author}{\bibfnamefont{J.}~\bibnamefont{de~Favereau}},
  \bibinfo{author}{\bibfnamefont{C.}~\bibnamefont{Delaere}},
  \bibinfo{author}{\bibfnamefont{P.}~\bibnamefont{Demin}},
  \bibinfo{author}{\bibfnamefont{A.}~\bibnamefont{Giammanco}},
  \bibinfo{author}{\bibfnamefont{V.}~\bibnamefont{Lemaître}},
  \bibinfo{author}{\bibfnamefont{A.}~\bibnamefont{Mertens}}, \bibnamefont{and}
  \bibinfo{author}{\bibfnamefont{M.}~\bibnamefont{Selvaggi}}
  (\bibinfo{collaboration}{DELPHES 3}), \bibinfo{journal}{JHEP}
  \textbf{\bibinfo{volume}{02}}, \bibinfo{pages}{057} (\bibinfo{year}{2014}),
  \eprint{1307.6346}.

\bibitem[{\citenamefont{Cepeda et~al.}(2019)}]{Cepeda:2019klc}
\bibinfo{author}{\bibfnamefont{M.}~\bibnamefont{Cepeda}} \bibnamefont{et~al.},
  \bibinfo{journal}{CERN Yellow Rep. Monogr.} \textbf{\bibinfo{volume}{7}},
  \bibinfo{pages}{221} (\bibinfo{year}{2019}), \eprint{1902.00134}.

\bibitem[{\citenamefont{Sirunyan et~al.}(2020)}]{Sirunyan:2020ztc}
\bibinfo{author}{\bibfnamefont{A.~M.} \bibnamefont{Sirunyan}}
  \bibnamefont{et~al.} (\bibinfo{collaboration}{CMS}), \bibinfo{journal}{Eur.
  Phys. J. C} \textbf{\bibinfo{volume}{80}}, \bibinfo{pages}{752}
  (\bibinfo{year}{2020}), \eprint{2001.10086}.

\bibitem[{\citenamefont{Esteban et~al.}(2019)\citenamefont{Esteban,
  Gonzalez-Garcia, Hernandez-Cabezudo, Maltoni, and Schwetz}}]{Esteban:2018azc}
\bibinfo{author}{\bibfnamefont{I.}~\bibnamefont{Esteban}},
  \bibinfo{author}{\bibfnamefont{M.}~\bibnamefont{Gonzalez-Garcia}},
  \bibinfo{author}{\bibfnamefont{A.}~\bibnamefont{Hernandez-Cabezudo}},
  \bibinfo{author}{\bibfnamefont{M.}~\bibnamefont{Maltoni}}, \bibnamefont{and}
  \bibinfo{author}{\bibfnamefont{T.}~\bibnamefont{Schwetz}},
  \bibinfo{journal}{JHEP} \textbf{\bibinfo{volume}{01}}, \bibinfo{pages}{106}
  (\bibinfo{year}{2019}), \eprint{1811.05487}.

\bibitem[{\citenamefont{Aad et~al.}(2021)}]{ATLAS:2021jol}
\bibinfo{author}{\bibfnamefont{G.}~\bibnamefont{Aad}} \bibnamefont{et~al.}
  (\bibinfo{collaboration}{ATLAS}), \bibinfo{journal}{JHEP}
  \textbf{\bibinfo{volume}{06}}, \bibinfo{pages}{146} (\bibinfo{year}{2021}),
  \eprint{2101.11961}.

\bibitem[{\citenamefont{Zee}(1980{\natexlab{b}})}]{ZEE1980389}
\bibinfo{author}{\bibfnamefont{A.}~\bibnamefont{Zee}},
  \bibinfo{journal}{Physics Letters B} \textbf{\bibinfo{volume}{93}},
  \bibinfo{pages}{389} (\bibinfo{year}{1980}{\natexlab{b}}), ISSN
  \bibinfo{issn}{0370-2693}.

\bibitem[{\citenamefont{Workman et~al.}(2022)}]{ParticleDataGroup:2022pth}
\bibinfo{author}{\bibfnamefont{R.~L.} \bibnamefont{Workman}}
  \bibnamefont{et~al.} (\bibinfo{collaboration}{Particle Data Group}),
  \bibinfo{journal}{PTEP} \textbf{\bibinfo{volume}{2022}},
  \bibinfo{pages}{083C01} (\bibinfo{year}{2022}).

\bibitem[{\citenamefont{Nebot et~al.}(2008)\citenamefont{Nebot, Oliver, Palao,
  and Santamaria}}]{Nebot:2007bc}
\bibinfo{author}{\bibfnamefont{M.}~\bibnamefont{Nebot}},
  \bibinfo{author}{\bibfnamefont{J.~F.} \bibnamefont{Oliver}},
  \bibinfo{author}{\bibfnamefont{D.}~\bibnamefont{Palao}}, \bibnamefont{and}
  \bibinfo{author}{\bibfnamefont{A.}~\bibnamefont{Santamaria}},
  \bibinfo{journal}{Phys. Rev. D} \textbf{\bibinfo{volume}{77}},
  \bibinfo{pages}{093013} (\bibinfo{year}{2008}), \eprint{0711.0483}.

\bibitem[{\citenamefont{Aad et~al.}(2015)}]{PhysRevD.91.052007}
\bibinfo{author}{\bibfnamefont{G.}~\bibnamefont{Aad}} \bibnamefont{et~al.}
  (\bibinfo{collaboration}{ATLAS Collaboration}), \bibinfo{journal}{Phys. Rev.
  D} \textbf{\bibinfo{volume}{91}}, \bibinfo{pages}{052007}
  (\bibinfo{year}{2015}).

\bibitem[{\citenamefont{Aaboud et~al.}(2017)}]{ATLAS:2017eqx}
\bibinfo{author}{\bibfnamefont{M.}~\bibnamefont{Aaboud}} \bibnamefont{et~al.}
  (\bibinfo{collaboration}{ATLAS}), \bibinfo{journal}{Phys. Rev. D}
  \textbf{\bibinfo{volume}{96}}, \bibinfo{pages}{052004}
  (\bibinfo{year}{2017}), \eprint{1703.09127}.

\bibitem[{\citenamefont{Aaboud et~al.}(2019{\natexlab{a}})}]{ATLAS:2018hbc}
\bibinfo{author}{\bibfnamefont{M.}~\bibnamefont{Aaboud}} \bibnamefont{et~al.}
  (\bibinfo{collaboration}{ATLAS}), \bibinfo{journal}{Phys. Lett. B}
  \textbf{\bibinfo{volume}{788}}, \bibinfo{pages}{316}
  (\bibinfo{year}{2019}{\natexlab{a}}), \eprint{1801.08769}.

\bibitem[{\citenamefont{{ATLAS Collaboration}}(2016)}]{ATLAS-CONF-2016-070}
\bibinfo{author}{\bibnamefont{{ATLAS Collaboration}}} (\bibinfo{year}{2016}),
  \eprint{ATLAS-CONF-2016-070}.

\bibitem[{\citenamefont{Aaboud et~al.}(2019{\natexlab{b}})}]{ATLAS:2019wdu}
\bibinfo{author}{\bibfnamefont{M.}~\bibnamefont{Aaboud}} \bibnamefont{et~al.}
  (\bibinfo{collaboration}{ATLAS}), \bibinfo{journal}{JHEP}
  \textbf{\bibinfo{volume}{05}}, \bibinfo{pages}{142}
  (\bibinfo{year}{2019}{\natexlab{b}}), \eprint{1903.01400}.

\bibitem[{\citenamefont{Aaboud et~al.}(2018{\natexlab{b}})}]{ATLAS:2018qto}
\bibinfo{author}{\bibfnamefont{M.}~\bibnamefont{Aaboud}} \bibnamefont{et~al.}
  (\bibinfo{collaboration}{ATLAS}), \bibinfo{journal}{Phys. Rev. Lett.}
  \textbf{\bibinfo{volume}{121}}, \bibinfo{pages}{081801}
  (\bibinfo{year}{2018}{\natexlab{b}}), \eprint{1804.03496}.

\bibitem[{\citenamefont{Babu and Mohapatra}(1995)}]{Babu:1995vh}
\bibinfo{author}{\bibfnamefont{K.~S.} \bibnamefont{Babu}} \bibnamefont{and}
  \bibinfo{author}{\bibfnamefont{R.~N.} \bibnamefont{Mohapatra}},
  \bibinfo{journal}{Phys. Rev. Lett.} \textbf{\bibinfo{volume}{75}},
  \bibinfo{pages}{2276} (\bibinfo{year}{1995}), \eprint{hep-ph/9506354}.

\bibitem[{\citenamefont{Antonelli
  et~al.}(2008)}]{FlaviaNetWorkingGrouponKaonDecays:2008hpm}
\bibinfo{author}{\bibfnamefont{M.}~\bibnamefont{Antonelli}}
  \bibnamefont{et~al.} (\bibinfo{collaboration}{FlaviaNet Working Group on Kaon
  Decays}) (\bibinfo{year}{2008}), \eprint{0801.1817}.

\bibitem[{\citenamefont{Mahmoudi et~al.}(2017)\citenamefont{Mahmoudi, Hurth,
  and Neshatpour}}]{Mahmoudi:2016mgr}
\bibinfo{author}{\bibfnamefont{F.}~\bibnamefont{Mahmoudi}},
  \bibinfo{author}{\bibfnamefont{T.}~\bibnamefont{Hurth}}, \bibnamefont{and}
  \bibinfo{author}{\bibfnamefont{S.}~\bibnamefont{Neshatpour}},
  \bibinfo{journal}{Nucl. Part. Phys. Proc.}
  \textbf{\bibinfo{volume}{285-286}}, \bibinfo{pages}{39}
  (\bibinfo{year}{2017}), \eprint{1611.05060}.

\bibitem[{\citenamefont{Amhis et~al.}(2022)}]{HFLAV:2022pwe}
\bibinfo{author}{\bibfnamefont{Y.}~\bibnamefont{Amhis}} \bibnamefont{et~al.}
  (\bibinfo{collaboration}{HFLAV}) (\bibinfo{year}{2022}), \eprint{2206.07501}.

\bibitem[{\citenamefont{Campbell and Maybury}(2005)}]{Campbell:2003ir}
\bibinfo{author}{\bibfnamefont{B.~A.} \bibnamefont{Campbell}} \bibnamefont{and}
  \bibinfo{author}{\bibfnamefont{D.~W.} \bibnamefont{Maybury}},
  \bibinfo{journal}{Nucl. Phys. B} \textbf{\bibinfo{volume}{709}},
  \bibinfo{pages}{419} (\bibinfo{year}{2005}), \eprint{hep-ph/0303046}.

\bibitem[{\citenamefont{Egana-Ugrinovic
  et~al.}(2021)\citenamefont{Egana-Ugrinovic, Homiller, and
  Meade}}]{Egana-Ugrinovic:2021uew}
\bibinfo{author}{\bibfnamefont{D.}~\bibnamefont{Egana-Ugrinovic}},
  \bibinfo{author}{\bibfnamefont{S.}~\bibnamefont{Homiller}}, \bibnamefont{and}
  \bibinfo{author}{\bibfnamefont{P.}~\bibnamefont{Meade}},
  \bibinfo{journal}{Phys. Rev. D} \textbf{\bibinfo{volume}{103}},
  \bibinfo{pages}{115005} (\bibinfo{year}{2021}), \eprint{2101.04119}.

\bibitem[{\citenamefont{Sirunyan et~al.}(2018)}]{CMS:2018yan}
\bibinfo{author}{\bibfnamefont{A.~M.} \bibnamefont{Sirunyan}}
  \bibnamefont{et~al.} (\bibinfo{collaboration}{CMS}), \bibinfo{journal}{JHEP}
  \textbf{\bibinfo{volume}{11}}, \bibinfo{pages}{151} (\bibinfo{year}{2018}),
  \eprint{1807.02048}.

\end{thebibliography}

\end{document}